\documentclass[]{aa}  

\usepackage{graphicx}
\usepackage{txfonts}
\usepackage{lipsum}
\usepackage{subcaption}         
\usepackage{lscape}             
\usepackage{placeins}           

\usepackage{natbib}
\usepackage{amsmath} 
\usepackage{amssymb}
\usepackage{mathtools}
\usepackage{microtype}
\usepackage[dvipsnames]{xcolor}
\definecolor{CiteRed}{RGB}{110, 0, 0}
\usepackage{multirow}
\usepackage{array}
\usepackage{makecell}
\usepackage{siunitx}
\usepackage{gensymb}
\usepackage{booktabs}
\usepackage{graphicx}
\usepackage{subcaption}
\usepackage{verbatim}
\usepackage{orcidlink}
\graphicspath{{fig/}}

\usepackage{hyperref}
\hypersetup{
    bookmarks,
    breaklinks,
    colorlinks,
    citecolor=CiteRed,  
    linkcolor=NavyBlue, 
    urlcolor=NavyBlue,  
    linktoc=page,
    pdftitle={A new $H_0$ measurement with SNe Requiem and Encore using Gravity.jl},
    pdfkeywords={Galaxies: clusters: general; Gravitational lensing: strong; Cosmology: observations},
    pdfauthor={Lorenzo Bazzanini},
    pdfcreator={\LaTeX}
}

\begin{document}

\title{A new $H_0$ measurement with SNe Requiem and Encore using Gravity.jl}

\author{
L.~Bazzanini\thanks{\email{lorenzo.bazzanini[at]unife[dot]it}}\inst{\ref{aff1},\ref{aff24},\ref{aff2},\ref{aff12}}\orcidlink{0000-0003-0727-0137}
\and G.~Di~Rosa\inst{\ref{aff1},\ref{aff2},\ref{aff12}}\orcidlink{0009-0001-9416-0923}
\and A.~Acebron\inst{\ref{aff6},\ref{aff7}}\orcidlink{0000-0003-3108-9039}
\and P.~Bergamini\inst{\ref{aff2}}\orcidlink{0000-0003-1383-9414}
\and M.~Lombardi\inst{\ref{aff4},\ref{aff2}}\orcidlink{0000-0002-3336-4965}
\and P.~Rosati\inst{\ref{aff1},\ref{aff2},\ref{aff12}}\orcidlink{0000-0002-6813-0632}
\and G.~Angora\inst{\ref{aff3},\ref{aff1}}\orcidlink{0000-0002-0316-6562}
\and G.~B.~Caminha\inst{\ref{aff17},\ref{aff18}}\orcidlink{0000-0001-6052-3274}
\and S.~Ertl\inst{\ref{aff18},\ref{aff17}}\orcidlink{0000-0002-5085-2143}
\and G.~Granata\inst{\ref{aff9}}\orcidlink{0000-0002-9512-3788}
\and A.~M.~Koekemoer\inst{\ref{aff14}}\orcidlink{0000-0002-6610-2048}
\and S.~H.~Suyu\inst{\ref{aff17},\ref{aff18}}\orcidlink{0000-0001-5568-6052}
\and J.~Pierel\inst{\ref{aff14}}\orcidlink{0000-0002-2361-7201}
\and A.~B.~Newman\inst{\ref{aff19}}\orcidlink{0000-0001-7769-8660}
\and C.~Grillo\inst{\ref{aff4},\ref{aff7}}\orcidlink{0000-0002-5926-7143}
\and S.~Schuldt\inst{\ref{aff15},\ref{aff16},\ref{aff7}}\orcidlink{0000-0003-2497-6334}
\and M.~Bulla\inst{\ref{aff1},\ref{aff12},\ref{aff13}}\orcidlink{0000-0002-8255-5127}
\and S.~Cha\inst{\ref{aff21}}\orcidlink{0000-0001-7148-6915}
\and M.~D'Addona\inst{\ref{aff3}}\orcidlink{0000-0003-3445-0483}
\and J.~M.~Diego\inst{\ref{aff6}}\orcidlink{0000-0001-9065-3926}
\and M.~Fogliardi\inst{\ref{aff1},\ref{aff2},\ref{aff12}}\orcidlink{0009-0006-4964-5311}
\and B.~L.~Frye\inst{\ref{aff20}}\orcidlink{0000-0003-1625-8009}
\and M.~J.~Jee\inst{\ref{aff21},\ref{aff22}}\orcidlink{0000-0002-5751-3697}
\and P.~S.~Kamieneski\inst{\ref{aff23}}\orcidlink{0000-0001-9394-6732}
\and M.~Meneghetti\inst{\ref{aff2},\ref{aff5}}\orcidlink{0000-0003-1225-7084}
\and A.~Mercurio\inst{\ref{aff3},\ref{aff8},\ref{aff10}}\orcidlink{0000-0001-9261-7849}
\and E.~Vanzella\inst{\ref{aff2}}\orcidlink{0000-0002-5057-135X}
}

\institute{Dipartimento di Fisica e Scienze della Terra, Universit\`a degli Studi di Ferrara, Via Giuseppe Saragat 1, 44122 Ferrara, Italy\label{aff1}
\and
INAF -- Istituto di Radioastronomia, via Gobetti 101, I-40129, Bologna, Italy\label{aff24}
\and
INAF -- Osservatorio di Astrofisica e Scienza dello Spazio di Bologna, Via Piero Gobetti 93/3, 40129 Bologna, Italy\label{aff2}
\and
INFN -- Sezione di Ferrara, via Saragat 1, I--44122, Ferrara, Italy\label{aff12}
\and
Instituto de F\'isica de Cantabria (CSIC-UC). Avda. Los Castros s/n. 39005 Santander, Spain\label{aff6}
\and
INAF -- IASF Milano, Via Alfonso Corti 12, 20133 Milano, Italy\label{aff7}
\and
Dipartimento di Fisica "Aldo Pontremoli", Universit\`a degli Studi di Milano, Via Celoria 16, 20133 Milano, Italy\label{aff4}
\and
INAF -- Osservatorio Astronomico di Capodimonte, Via Moiariello 16, 80131 Napoli, Italy\label{aff3}
\and
Technical University of Munich, TUM School of Natural Sciences, James-Franck-Straße 1, 85748 Garching, Germany\label{aff17}
\and
Max-Planck-Institut für Astrophysik, Karl-Schwarzschild Straße 1, 85748 Garching, Germany\label{aff18}
\and
Institute of Cosmology and Gravitation, University of Portsmouth, Burnaby Road, Portsmouth PO1 3FX, UK\label{aff9}
\and
Space Telescope Science Institute, 3700 San Martin Drive, Baltimore, MD 21218, USA\label{aff14}
\and
Observatories, Carnegie Science, Pasadena, CA 91101, USA\label{aff19}
\and
Finnish Centre for Astronomy with ESO (FINCA), University of Turku, FI-20014 Turku, Finland\label{aff15}
\and
Department of Physics, P.O. Box 64, University of Helsinki, FI-00014 Helsinki, Finland\label{aff16}
\and
INAF -- Osservatorio Astronomico d’Abruzzo, via Mentore Maggini snc, 64100 Teramo, Italy\label{aff13}
\and
Department of Astronomy/Steward Observatory, University of Arizona, 933 N. Cherry Avenue, Tucson, AZ 85721, USA\label{aff20}
\and 
Department of Astronomy, Yonsei University, 50 Yonsei-ro, Seoul
03722, Korea\label{aff21}
\and 
Department of Physics and Astronomy, University of California,
Davis, One Shields Avenue, Davis, CA 95616, USA\label{aff22}
\and
Department of Physics and Astronomy, Chalmers University of Technology, SE-412 96 Gothenburg, Sweden\label{aff23}
\and
INFN -- Sezione di Bologna, Viale Berti Pichat 6/2, 40127 Bologna, Italy\label{aff5}
\and
Universita di Salerno, Dipartimento di Fisica "E.R. Caianiello", Via Giovanni Paolo II 132, I-84084 Fisciano (SA), Italy\label{aff8}
\and
INFN -- Gruppo Collegato di Salerno, Sezione di Napoli, Dipartimento di Fisica "E.R. Caianiello", Universita di Salerno, via Giovanni Paolo II, 132 - I-84084 Fisciano (SA), Italy\label{aff10}
}

\date{Received xxx; accepted xxx}


\abstract{
We present a strong-lensing analysis of the galaxy cluster MACS~J0138.0$-$2155 ($z_\mathrm{l}=0.336$), the first known cluster lens discovered to host two distinct multiply imaged Type~Ia supernovae (SNe): SN~Requiem and SN~Encore. 
Both transients are located in the massive, multiply imaged red galaxy MRG-M0138 at $z_\mathrm{s}=1.949$. 
The projected total mass distribution of this cluster has been investigated with several independent lens models (Suyu et al. 2026, A\&A, 708, A291; Pierel et al. 2026, ApJ, 998, 219), using a sample of $23$ spectroscopically confirmed multiple images from eight background sources spanning $0.767 < z < 3.420$, identified from Hubble Space Telescope and James Webb Space Telescope imaging, and VLT/MUSE spectroscopy. 
In this work, we develop a new lens model with the Bayesian parametric lens-modelling framework \textsc{Gravity.jl}, using the same strong-lensing dataset. 
Under a fiducial flat $\Lambda$CDM cosmology, our reference mass model accurately reproduces the observed image positions, with an image-plane root-mean-square positional residual of $\Delta_{\rm rms}=0.15''$. 
Assuming $H_0=70\,\mathrm{km}\,\mathrm{s}^{-1}\,\mathrm{Mpc}^{-1}$, we predict the future reappearances of highly delayed SN counter-images, finding $\Delta t_{\rm 1d,1a}=3122^{+75}_{-70}$ days ($\approx$ April--August 2032) for SN~Encore and $\Delta t_{\rm 2d,2a}=3902^{+67}_{-72}$ days ($\approx$ January--May 2027) for SN~Requiem. 
By allowing the Hubble constant to vary and including the measured relative time delays of both SN~Encore and SN~Requiem, together with their statistical uncertainties, as observables, we infer the value of $H_0$ jointly with the lens-model parameters. 
We performed the cosmographic analysis with three alternative parametrisations of the large-scale cluster halo and combine the resulting $H_0$ posteriors through Bayesian model averaging, using the corresponding Bayesian evidences as model weights. 
The resulting model-averaged constraint is $H_0=69.2^{+9.6}_{-8.0}\,\mathrm{km}\,\mathrm{s}^{-1}\,\mathrm{Mpc}^{-1}$, consistent with the value inferred from independent lens models of this cluster. 
The current uncertainty remains dominated by the relatively large fractional uncertainties of the measured SN time delays ($\gtrsim10\%$), while the scatter among the mass parametrisations considered here is subdominant. 
The forthcoming long-baseline reappearance of SN~Requiem offers an immediate opportunity to improve the statistical precision on the Hubble constant, provided that lens-model systematics are sufficiently controlled. 
Together, these results establish MACS~J0138.0$-$2155 as a premier anchor for high-precision cluster-scale time-delay cosmography with multiply imaged SNe.  
}

\keywords{
    Galaxies: clusters: general --
    Gravitational lensing: strong --
    Cosmology: observations 
}

\maketitle
\nolinenumbers 


\section{Introduction}
\label{sec:introduction}

Strong gravitational lensing (SL) of variable or transient background sources provides a direct geometrical probe of the cosmological distance scale through time-delay cosmography~\citep[TDC;][]{refsdal64, narayan96, treu16}. 
Because the relative arrival times of multiple images scale to leading order as the inverse of the Hubble constant ($H_0$), TDC offers an independent and complementary avenue for studying the current tension between early- and late-Universe measurements of $H_0$~\citep[e.g.][]{divalentino21, verde24}. 
While TDC has historically focused on galaxy-scale multiply imaged quasars~\citep[QSOs; e.g.][]{suyu10, wong20, millon20, birrer25}, cosmological inferences from galaxy-scale lenses can be limited by systematic uncertainties associated with the mass-sheet degeneracy, assumptions on the lens radial total mass profile, microlensing, and line-of-sight structures~\citep[e.g.][]{falco85, saha00, schneider13, xu16, shajib23}.

Galaxy clusters provide a rich but complex alternative regime for TDC. 
Their extended critical curves generate numerous multiple images across wide redshift baselines that tightly constrain the total projected mass distribution, despite challenges in modelling cluster-scale dark matter haloes, galaxy substructures, and line-of-sight perturbers~\citep[e.g.][]{bradac04, grillo15, meneghetti17, cha22, bergamini23a}. 
Within this regime, strongly lensed supernovae (SNe) have emerged as particularly valuable cosmographic probes~\citep[e.g.][]{suyu20, grillo24, pascale25, suyu26}. 
Unlike QSOs, Type~Ia SNe have predictable light curves, standardisable luminosities, and characteristic colour evolution~\citep[e.g.][]{pierel26}. 
These properties enable precise time-delay and magnification measurements, while their transient nature allows late-time template imaging to separate the SN signal from the host-galaxy emission~\citep[e.g.][]{rodney21, arendse24}. 
Following the milestone discovery and model validation of SN~Refsdal behind MACS~J1149.5$+$2223~\citep{kelly15, kelly16a, kelly23, grillo16, grillo18, grillo20, grillo24}, expanding the sample of cluster-lensed SNe has become a key goal for current and upcoming surveys~\citep[e.g.][]{oguri10, goldstein17, goldstein19, wojtak19, frye24, bronikowski25, johansson25, coulter26, dhanasingham26, zitrin26, kamieneski26}. 

The massive galaxy cluster MACS~J0138.0$-$2155~\citep[$z_{\mathrm{l}} = 0.336$;][]{ebeling01}, hereafter M0138, is a premier laboratory for cluster TDC. 
It lenses the massive red galaxy MRG-M0138 at $z_{\mathrm{s}} = 1.949$~\citep{newman18, newman26}, which uniquely hosted two separate multiply imaged Type~Ia SNe: SN~Requiem~\citep{rodney21} and SN~Encore~\citep{dhawan24, pierel24, pierel26}\footnote{SN~Encore was spectroscopically classified as a Type~Ia SN by~\citet{dhawan24}. SN~Requiem was not spectroscopically confirmed, but its photometric properties and the massive, passive nature of its host galaxy strongly favour a Type~Ia classification~\citep{rodney21}.}.  
This is the first known lens cluster in which two different multiply imaged SNe have been observed in the same host galaxy, providing independent constraints on the gravitational potential. 
Recent deep imaging from the Hubble Space Telescope (HST) and the James Webb Space Telescope~\citep[JWST;][]{gardner06}, combined with spectroscopy from the Multi Unit Spectroscopic Explorer~\citep[MUSE;][]{bacon14} on the Very Large Telescope, has established a robust ``gold'' sample of $23$ spectroscopically confirmed multiple images from eight background sources~\citep{granata25, ertl25}. 
Building upon this extensive spectro-photometric dataset and existing SL models of this cluster~\citep[e.g.][]{acebron25, ertl25, suyu26}, this system provides an unprecedented baseline for breaking modelling degeneracies, cross-validating independent model predictions, and ultimately translating the long predicted time delays of upcoming SNe counter-images into a robust, independent measurement of the value of $H_0$. 
In this context, \citet{suyu26} recently compared several independent lens models of M0138, using them to predict the future reappearances of SN~Requiem and SN~Encore and, together with the measured SN~Encore time delay from \citet{pierel26}, to infer the value of $H_0$. 

In this work, we present a new SL and cosmological analysis of M0138 using the novel Bayesian parametric lens-modelling framework \textsc{Gravity.jl}~\citep{lombardi24}, building upon the analysis by~\citet{acebron25}. 
Constrained by the spectroscopic multiple-image sample, we first reconstruct the cluster total mass distribution under a fixed fiducial cosmology to predict the configurations, time delays, and magnifications of the upcoming counter-images for both SN~Requiem and SN~Encore. 
We then include the measured relative time delays between multiple images of SN~Encore~\citep{pierel26} and SN~Requiem~\citep{rodney21} as additional observables in the SL likelihood, and infer $H_0$ jointly with the parameters describing the total mass distribution of the lens within a flat $\Lambda$CDM framework.

Throughout this work, we adopt a fiducial flat $\Lambda$CDM cosmology defined by $H_0=70\,\mathrm{km}\,\mathrm{s}^{-1}\, \mathrm{Mpc}^{-1}$, $\Omega_\mathrm{m}=0.3$, and $\Omega_{\Lambda}=0.7$, unless otherwise stated. 
In this cosmology, an angular separation of $1''$ corresponds to a physical scale of $4.81$ kpc at the lens redshift, $z_{\mathrm{l}} = 0.336$. 
All magnitudes are reported in the absolute bolometric (AB) system~\citep{oke74}. 
Quoted uncertainties represent the $68\%$ Bayesian credible interval.

The paper is organised as follows. 
Section~\ref{sec:data} summarises the imaging and spectroscopic data. 
Section~\ref{sec:modeling} describes the SL constraints and total mass parametrisation of M0138. 
Section~\ref{sec:results} presents the fixed-cosmology model, the surface-brightness (SB) reconstruction of MRG-M0138, the SNe predictions, and the joint $H_0$ inference. 
Section~\ref{sec:conclusions} summarises our conclusions and outlines future prospects.


\section{Observational data}
\label{sec:data}

This section summarises the HST and JWST multi-band imaging, and the MUSE and JWST spectroscopic observations used to construct the SL mass model of M0138. 
A detailed description of the imaging and spectroscopic datasets is provided by~\citet{pierel24, ertl25, granata25}. 

\subsection{HST and JWST imaging}

M0138 was observed with HST in 2016 (ID 14496; PI: Newman), 2019 (ID 15663; PI: Akhshik), and again between December 2023 and January 2024 (ID 16264; PI: Pierel). 
SN Requiem was identified by~\citet{rodney21} in the 2016 HST imaging, after the corresponding multiple images were found to be absent in the 2019 observations. 
The first JWST observations of M0138 were obtained in November 2023 (ID 2345; PI: Newman), leading to the discovery of SN Encore~\citep{pierel24}. 
This discovery motivated follow-up observations in December 2023 and January 2024 with HST (ID 16264; PI: Pierel) and JWST (ID 6549; PI: Pierel), aimed at securing the light curves of the multiple images of SN Encore. 

\begin{figure*}
    \centering
     
     \includegraphics[width=0.99\textwidth]{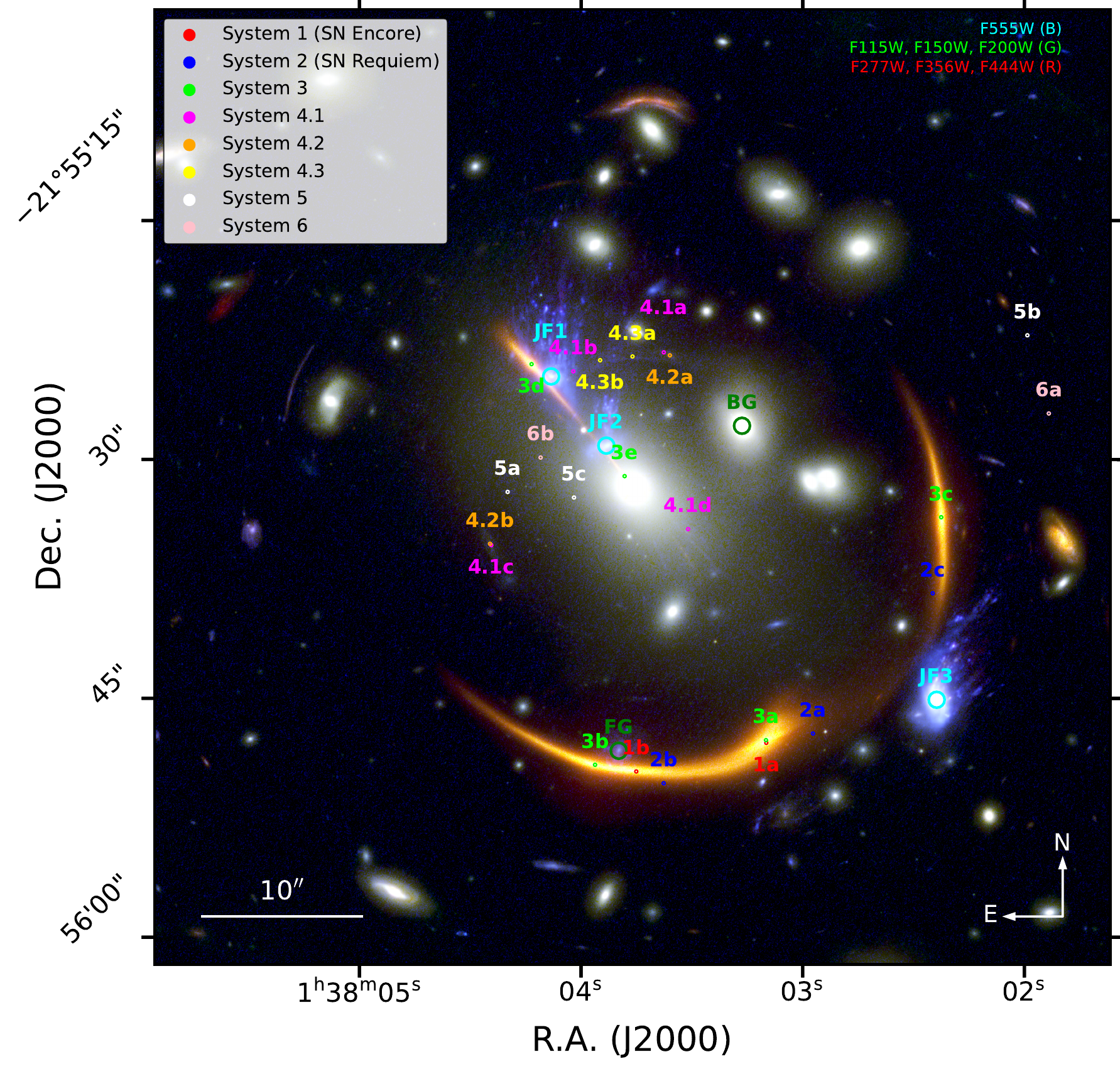}
     \caption{M0138 observed with the JWST and the HST. The colour composite is constructed from the following filter combinations: F555W (HST) in blue, F115W$+$F150W$+$F200W (JWST) in green, and F277W$+$F356W$+$F444W (JWST) in red. The locations of the ``gold'' multiple-image systems are marked with circles~\citep{ertl25, suyu26}. SN Encore corresponds to System 1, while SN Requiem corresponds to System 2. The foreground (FG) and background (BG) galaxies are indicated by green circles. The three jellyfish galaxies are marked with cyan circles and labelled JF1, JF2, and JF3.} 
      \label{fig:m0138}
\end{figure*}

\subsection{MUSE and JWST spectroscopy}

M0138 was observed with MUSE as part of two observing programmes. 
The first programme (ID 0103.A-0777; PI: Edge) targeted the central $1' \times 1'$ field of view of the cluster in September 2019, for a total integration time of $\approx49$ minutes. 
Following the discovery of SN Encore, an additional $\approx2.9$ hours of MUSE observations of the same region were obtained in December 2023 (ID 110.23PS; PI: Suyu). 
The average seeing across the observations was $\approx0.8''$. 
The data from both programmes were reduced and analysed by~\citet{granata25}, who constructed a spectroscopic catalogue containing $107$ reliable redshift measurements of extragalactic sources within the cluster field. 
This sample includes both cluster members and multiply imaged background galaxies, which are used to build the SL mass model. 

SN Encore and its host galaxy were also observed with the Near InfraRed Spectrograph integral-field unit aboard JWST, targeting the multiple images 1a and 3a (see Fig.~\ref{fig:m0138}; ID 2345; PI: Newman). 
Spectroscopic observations obtained at two different epochs allowed~\citet{dhawan24} to classify SN Encore, at $z_{\rm s}=1.949$, as a normal Type Ia SN whose spectral characteristics closely resemble those of nearby Type Ia SNe. 

%



\section{Strong-lensing modelling}
\label{sec:modeling}

In this section, we describe the SL mass parametrisations of M0138 adopted in our work. We performed the analysis using the newly developed lens-modelling software \textsc{Gravity.jl}~\citep{lombardi24}, which implements a full Bayesian framework for the parametric reconstruction of galaxy-cluster total mass distributions. 

The posterior distributions of the model parameters were sampled using the non-reversible parallel-tempering algorithm implemented in \texttt{Pigeons.jl}~\citep{syed19, surjanovic22, surjanovic23}, and the affine-invariant ensemble sampler \texttt{emcee}~\citep{goodman10, foreman-mackey13}. 
All lens-model optimisations were performed in the ``simplified image-plane'' configuration; we refer the reader to~\citet{lombardi24} for details of this implementation.

\subsection{Cluster members}

The cluster-member catalogue adopted here is taken from the spectroscopic analysis of~\citet{granata25}, complemented by the photometric study of~\citet{ertl25}. 
It includes $84$ cluster members with $\mathrm{F160W} < 24$ within the field of view of M0138, $50$ of which are spectroscopically confirmed. 
From these secure spectroscopic members, \citet{granata25} derived a mean cluster redshift of $z_\mathrm{l} = 0.336$.
Among the $84$ cluster members, three are classified as jellyfish galaxies~\citep[e.g.][]{ebeling14, gibson25}. 
These galaxies, labelled JF1, JF2, and JF3 in Fig.~\ref{fig:m0138}, were modelled separately in the SL model with individually optimised galaxy-scale mass components and were therefore excluded from the cluster-member scaling relation. 

\citet{granata25} measured central stellar velocity dispersion values for $14$ early-type cluster galaxies with sufficiently high spectral signal-to-noise ratio, $\langle\mathrm{S/N}\rangle>10$, down to $\mathrm{F160W}\simeq21$, including the brightest cluster galaxy (BCG). 
These measurements allowed us to calibrate the Faber--Jackson relation~\citep{faber76} for the cluster members~\citep[e.g.][]{bergamini19}, linking their luminosities to their velocity dispersions and, consequently, to their total masses. 
The photometric and velocity-dispersion catalogues are publicly released by~\citet{ertl25} and~\citet{granata25}, respectively.

\subsection{Multiple image systems}

As lensing constraints, we adopted the ``gold'' multiple-image catalogue compiled from the analyses of~\citet{granata25}, \citet{ertl25}, and~\citet{suyu26}.  
This sample consists of $23$ spectroscopically confirmed images associated with eight distinct background sources, spanning the redshift interval $0.767<z<3.420$ (see Table~4 in~\citealt{ertl25}). 
Figure~\ref{fig:m0138} shows the JWST+HST colour image of M0138, with the positions of SN Encore (System 1), SN Requiem (System 2), and the other multiple-image systems used in the SL model. 
For all systems, we assigned elliptical positional uncertainties that reflect the observed spatial extent of each image in the dataset from which the positions are measured~\citep{ertl25}. 

\subsection{Time-delay measurements}

We summarise the time-delay measurements available for the multiple images of SN Requiem~\citep{rodney21} and SN Encore~\citep{pierel26}. 

\subsubsection{SN Requiem}

\citet{rodney21} measured the relative time delays between multiple images 2a, 2b, and 2c of SN Requiem using the predictable colour and luminosity time evolution of Type Ia SNe~\citep{kasen07,larison25}, although the Type~Ia classification was not secure from the available data alone. 
Time delays are commonly measured by fitting the light curves of the individual SN images and comparing their epochs of peak brightness~\citep{pierel19,dhawan20}. 
For SN Requiem, however, only a single epoch of photometry was available from the archival HST observations, preventing a standard light-curve fit and introducing strong degeneracies among the SN model parameters. 
Because gravitational lensing is achromatic, the observed colours of the multiple images are not altered by the lens, provided that differential extinction and chromatic microlensing are negligible~\citep{goldstein18}. 
The colour evolution can therefore be used to constrain the relative phase, and hence the relative arrival time, of each image. 
To estimate these phases, \citet{rodney21} used the SuperNova Time Delays code~\citep[SNTD;][]{pierel19} to reconstruct the intrinsic colour curve of SN Requiem with a SALT2 model template~\citep{guy07}. 
The method simultaneously fits the ages of all observed images while varying the SN model parameters. 
The resulting measured time delays, defined relative to image 2b, are $\Delta t_{\mathrm{2a,2b}} = 116.2^{+29.0}_{-29.9}$ days for image 2a, and $\Delta t_{\mathrm{2c,2b}} = 1.3^{+32.5}_{-32.9}$ days for image 2c (cf.~Extended Data Fig.~5 in \citealt{rodney21}).\footnote{In~\citet{suyu26}, the SN Requiem time delays are instead defined relative to image 2a.} 

\subsubsection{SN Encore}

\citet{pierel26} measured the multi-band JWST photometry of SN~Encore multiple images 1a and 1b, while the photometry of image 1c could not be reliably extracted from the available data. 
Using these flux measurements, they inferred a time delay of image 1b relative to image 1a of $\Delta t_{\mathrm{1b,1a}} = -39.8^{+3.9}_{-3.3}$ days. 
This measurement was obtained by fitting Type Ia SN template light curves from \textsc{BayesSN}~\citep{mandel22, ward23, grayling24} with two independent fitting frameworks, \textsc{Glimpse}~\citep{hayes24, hayes26} and SNTD~\citep{pierel19}. 
The quoted uncertainties include the effect of microlensing, modelled through simulations adopting four different Type Ia SN progenitor scenarios~\citep{huber21}. 
The resulting fractional time-delay uncertainty is approximately $10\%$, dominated by the photometric precision. 

\subsection{Total mass parametrisation}

Following~\citet{acebron25}, we parametrised the total gravitational potential of the lens, $\phi_{\mathrm{tot}}$, as the superposition of five components: 
(1) one large-scale smooth halo ($N_{\rm h}=1$) describing the cluster dark-matter distribution; 
(2) $N_{\rm gal}=81$ cluster-member galaxies described by luminosity-based scaling relations; 
(3) $N_{\rm LoS}=2$ individually modelled line-of-sight galaxies at redshifts different from that of the cluster, FG and BG; 
(4) $N_{\rm pert}=3$ individually modelled cluster members, corresponding to the three jellyfish galaxies; and 
(5) an external-shear term accounting for residual perturbations from mass structures not explicitly included in the lens model, both in the cluster environment and along the line of sight. 
Accordingly, the total gravitational potential is written as
\begin{equation}
    \phi_{\mathrm{tot}} = 
    \sum_{i=1}^{N_{h}=1}              \phi_i^{\mathrm{halo}}        +
    \sum_{j=1}^{N_{\mathrm{gal}}=81}  \phi_j^{\mathrm{gal}}         +
    \sum_{k=1}^{N_{\mathrm{LoS}}=2}   \phi_k^{\mathrm{gal,\, LoS}}  +
    \sum_{u=1}^{N_{\mathrm{pert}}=3}  \phi_u^{\mathrm{gal,\, pert}} +
    \phi_{\gamma} \, . 
\end{equation}

To reconstruct the total mass distribution of the cluster, we optimised the mass model by searching for the set of parameters $\vec{\xi}$ minimising the positional $\chi^2$ goodness-of-fit statistic on the image plane~\citep[e.g.][]{jullo07}
\begin{equation} 
    \chi^2_{\mathrm{pos}}(\vec{\xi}) =  \sum_{j=1}^{N_{\mathrm{fam}}}\sum_{i=1}^{N_{\mathrm{im},j}} \left(\frac{\|\vec{\vartheta}_{i,j}^{\mathrm{obs}}-\vec{\vartheta}_{i,j}^{\mathrm{pred}}(\vec{\xi})\|}{\Delta \vartheta{i,j}}\right)^2 \, ,
    \label{eq:chi2pos}
\end{equation}
where $N_{\mathrm{fam}}$ denotes the number of independent source families, and $N_{\mathrm{im},j}$ is the number of multiple images belonging to the $j$-th family. Here, $\vec{\vartheta}_{i,j}^{\mathrm{obs}}$ and $\vec{\vartheta}_{i,j}^{\mathrm{pred}}(\vec{\xi})$ represent the observed and model-predicted positions of the $i$-th image in family $j$, respectively, while $\Delta \vartheta_{i,j}$ accounts for the uncertainty on the observed image positions. 
Equation~\eqref{eq:chi2pos} describes only the positional part of the likelihood, which is used for the fixed-cosmology reference model; the measured SNe time delays do not enter this model. 
They are included instead in the models where $H_0$ is allowed to vary, entering the likelihood as additional observables through a separate time-delay contribution, $\chi^2_{\rm td}$. 
For these models, the total goodness of fit is therefore written as $\chi^2_{\rm tot} = \chi^2_{\rm pos} + \chi^2_{\rm td}$. 

To quantify the overall goodness of fit, we used the root mean square (rms) deviation between the observed and predicted positions of the multiple images 
\begin{equation}
    \Delta_\mathrm{rms}=
    \sqrt{
    \frac{1}{N_\mathrm{im}^\mathrm{tot}}
    \sum_{j=1}^{N_\mathrm{fam}}
    \sum_{i=1}^{N_{\mathrm{im},j}}
    \left\|
    \vec{\vartheta}_{i,j}^\mathrm{obs}
    -
    \vec{\vartheta}_{i,j}^\mathrm{pred}(\boldsymbol{\xi})
    \right\|^2
    } ,
\end{equation}
where $N_{\mathrm{im}}^{\mathrm{tot}}$ is the total number of multiple images. 

To compare the relative performance of the different models presented here, we use the Bayesian log-evidence, $\log Z$~\citep[][]{ivezic20}. 
The Bayesian Information Criterion~\citep[BIC;][]{schwarz78} is also computed (see Table~\ref{tab:bestmodel}) to compare our models with those in~\citet{suyu26}. 
Larger values of $\log Z$ and smaller values of the BIC indicate models which better describe the data~\citep[][]{ivezic20}.
%

\subsubsection{Dark matter mass distribution}

We modelled the large-scale halo, which dominates the cluster dark-matter component, using three alternative parametrisations: a Softened Power-law Elliptical Mass Distribution~\citep[SPEMD;][]{kormann94, barkana98}, a Navarro--Frenk--White profile~\citep[NFW;][]{navarro97,keeton01} profile, and a Non-singular Isothermal Ellipsoid~\citep[NIE;][]{keeton01}. 
In this section, we focus on the SPEMD model, which is favoured by the Bayesian evidence, while the comparison with the other parametrisations is discussed in Sect.~\ref{subsec:free}. 

The SPEMD profile is described by seven free parameters: the lens centroid, $(x,y)$, the axis ratio, $q$, the position angle, $\theta$, the Einstein radius (for the equivalent singular profile), $r_\mathrm{e}$ , the core radius, $r_{\mathrm{core}}$, and the positive slope of the mass profile $\alpha = 3-\gamma$ ($\gamma$ being the slope of the mass density profile $\rho \sim r^{-\gamma}$). 
The axis ratio is defined as the ratio of the semi-minor to semi-major axis, and the position angle is measured clockwise from North. 
In \textsc{Gravity.jl}, the SPEMD surface mass density $\kappa$ has elliptical iso-density contours~\citep{keeton01}, 
\begin{equation}
    \kappa_{\rm SPEMD}(x,y) = \frac{\alpha}{2} r_\mathrm{e}^{2-\alpha}
    \left(r_{\mathrm{core}}^2 + qx^2 + y^2/q\right)^{\alpha/2-1}\, .
\end{equation}

\subsubsection{Galaxy-scale mass distribution}
\label{sec:scalingrel}

The cluster-member galaxies were included in the lens model as sub-haloes described by singular, circular dual Pseudo-Isothermal Ellipsoid profiles~\citep[dPIE;][]{limousin05, eliasdottir07}. 
To significantly reduce the dimensionality of the parameter space, we scaled the total mass properties of the cluster-member galaxies with their near-infrared luminosities, which serve as a robust proxy for stellar mass~\cite[e.g.][]{grillo15}. 

Following~\citet{brainerd96, natarajan97, jullo07}, the velocity dispersion, $\sigma_{i}$, and truncation radius, $r_{\mathrm{cut},i}$, of the $i$-th cluster member are related to its luminosity, $L_i$, through power laws as 
\begin{equation} 
    \sigma_{i}= \sigma_{0}^{\mathrm{ref}} \left(\frac{L_i}{L_0} \right)^{\alpha} \,, \quad r_{\mathrm{cut},i}= r_{\mathrm{cut},0} \left( \frac{L_i}{L_0} \right)^{\beta_{\mathrm{cut}}} 
    \label{eq:scaling-rel} \, , 
\end{equation}
where $L_0$ is the luminosity of a reference galaxy, chosen to be the BCG, with $m_{\mathrm{F160W}}=15.30$. 

\begin{figure}
    \centering
    \includegraphics[width=0.99\linewidth]{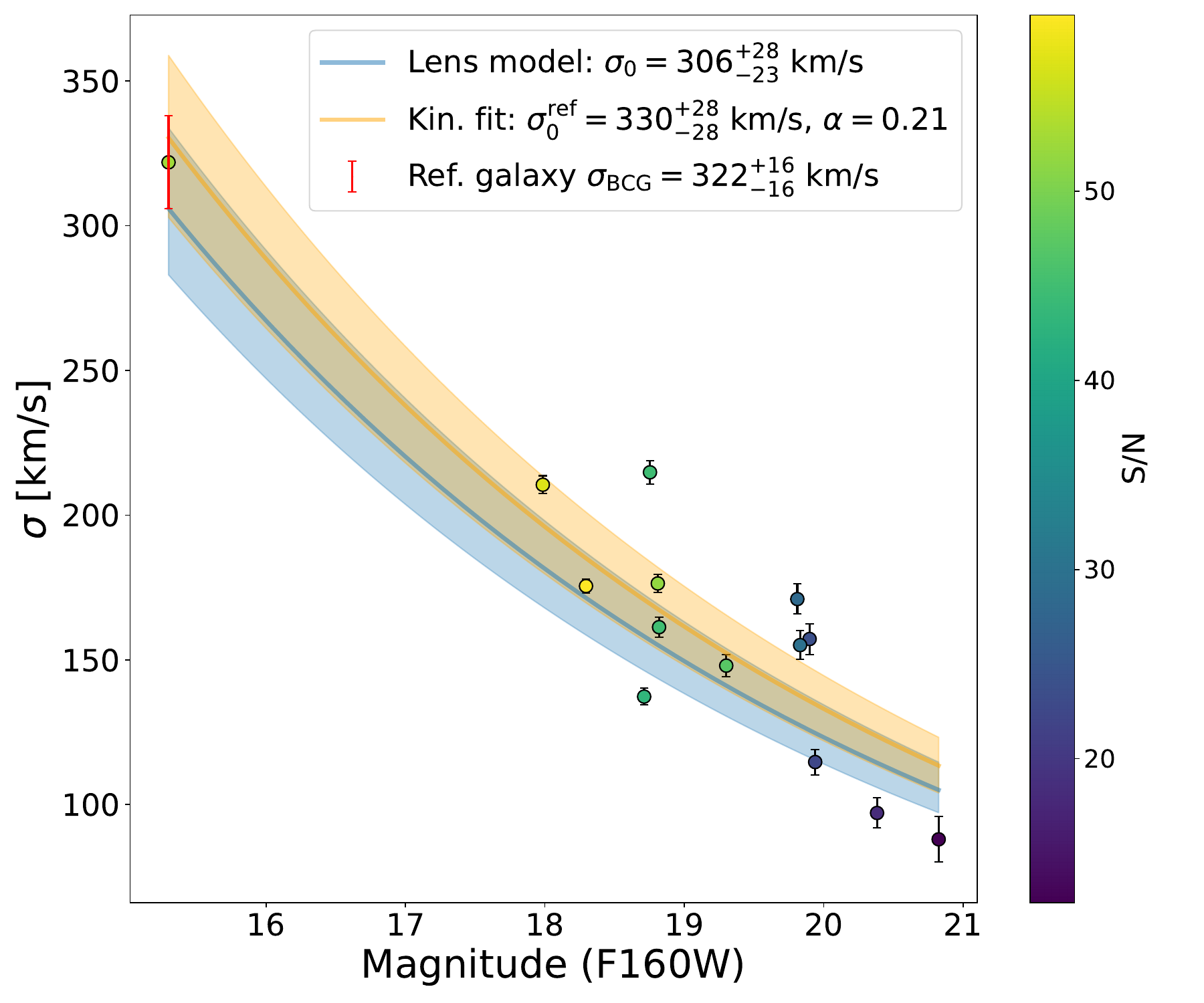}
    \caption{Calibration of the Faber--Jackson relation for cluster members from~\citet{acebron25}. The circles show the measured stellar velocity dispersions of 14 bright cluster galaxies as a function of their total HST F160W magnitude. Points are colour-coded by the average spectral signal-to-noise ratio reported by~\citet{granata25}. The reference galaxy used in the scaling relation is the BCG ($\sigma_\mathrm{BCG} = 322^{+16}_{-16}\,\mathrm{km}\,\mathrm{s}^{-1}$, $m_{\mathrm{F160W}}=15.30$). The orange line and shaded region show the median $\sigma_0$--$\mathrm{F160W}$ relation and its $68\%$ confidence interval derived from the stellar kinematic measurements, following the procedure described in Sect.~\ref{sec:scalingrel}. The blue line and shaded region show the corresponding relation predicted by the SL model.}
    \label{fig:scalingrel}
\end{figure}

By fixing the slopes $\alpha$ and $\beta_{\mathrm{cut}}$, the full cluster-member population is described in the lens model by only two free parameters: the normalisation of the velocity-dispersion scaling relation, $\sigma_{0}^{\mathrm{ref}}$, and the truncation-radius normalisation, $r_{\mathrm{cut},0}$. 
Following~\citet{bergamini19}, we used the central stellar velocity dispersions measured by~\citet{granata25} to calibrate the $\sigma_0$--$L$ relation by fitting $\sigma_{0}^{\mathrm{ref}}$ and $\alpha$ (see Fig.~\ref{fig:scalingrel}).  
We adopted the best-fitting values reported in Table~1 of~\citet{acebron25}, quoted in Fig.~\ref{fig:scalingrel}, and used the corresponding normalisation of the velocity-dispersion scaling relation as a Gaussian prior on $\sigma_{0}$ in the SL model. 
Assuming a non-constant total mass-to-light ratio scaling ${M_i}/{L_i} \propto L_i^{\gamma}$ for the cluster member population, the slopes in Eqs.~\eqref{eq:scaling-rel} satisfy $\beta_{\mathrm{cut}}=1-2\alpha+\gamma$. 
Thus, adopting a total mass-to-light ratio slope of $\gamma=0.2$, which is consistent with the fundamental plane~\citep{faber87, bender92}, \citet{acebron25} obtained a value of $\beta_{\mathrm{cut}}=0.78$.

\subsubsection{Line-of-sight galaxies and perturbers}

The model also includes five individually optimised perturbers, parametrised as singular spherical dPIE profiles. 
Three of them are the jellyfish cluster members JF1, JF2, and JF3, which are modelled outside the scaling relations. 
We also used a multi-plane framework for two spectroscopic line-of-sight galaxies identified by~\citet{granata25}: the foreground spiral galaxy near the southern giant arc, at $z=0.309$ and labelled FG, and the background galaxy near the BCG, at $z=0.371$ and labelled BG. 
For the BG galaxy, we adopted in the SL model a Gaussian prior on the velocity dispersion, $\sigma_\mathrm{bg}$, based on the stellar-kinematic measurement reported by~\citet{granata25}, as listed in Table~\ref{tab:modelparams}. 
These objects are shown in Fig.~\ref{fig:m0138}. 

\subsubsection{External shear term}

We included an external shear term to account for residual perturbations from non modelled mass in the cluster environment, line-of-sight mass structures, and deviations from the adopted parametric description of the main cluster halo~\citep[e.g.][]{lagattuta19, acebron22}. 

The external shear was modelled in polar coordinates $(r,\theta)$ following the standard parametrisation by~\cite[e.g.][]{keeton01}. The corresponding lensing potential is given by\footnote{Changing the arbitrary origin used to write the external-shear potential only adds a constant term and a term linear in the image-plane coordinates to the lensing potential $\psi$; neither affects the SL observables~\citep{keeton01}.} 
\begin{equation}
    \psi_{\gamma}(r, \theta) = \gamma \frac{r^2}{2}  \cos 2(\theta - \theta_\gamma) \, .
\end{equation}
Here, $\gamma$ represents the shear magnitude, and $\theta_\gamma$ defines the shear position angle, measured clockwise from the North axis.


\section{Results and discussion}
\label{sec:results}

In this section, we present the results of our SL analysis of M0138. 
We first describe in Sect.~\ref{subsec:fixed} the reference model obtained at fixed cosmology. 
In Sect.~\ref{subsec:sb}, we use this model to reconstruct the SB distribution of the multiply imaged host galaxy MRG-M0138 as a posterior predictive check. 
We then use the fixed-cosmology posterior in Sect.~\ref{subsec:predictions} to predict the image configurations, time delays, and magnifications of SN Encore and SN Requiem, and to compare our predictions with those from the independent lens models presented by~\citet{suyu26}. 
Finally, in Sect.~\ref{subsec:free}, we allow $H_0$ to vary in the SL model and infer its posterior distribution using the measured SNe time delays.

\subsection{Mass model with fixed cosmology}
\label{subsec:fixed}

We first optimised the mass model, hereafter referred to as \texttt{H0\_fixed\_SPEMD}, while keeping the cosmological parameters fixed to the reference values reported in Sect.~\ref{sec:introduction}. 
The posterior distribution was sampled with the non-reversible parallel-tempering MCMC algorithm implemented in \texttt{Pigeons.jl}. 
We evolve $60$ chains over $16$ rounds until convergence is reached, corresponding to a total wall-clock time of approximately $40$ hours on a $10$-core workstation. 
We assessed convergence using the potential scale-reduction factor, $\hat{R}$~\citep{gelman92}, finding values well below the commonly adopted threshold $\hat{R}<1.1$~\citep{gelman13}. 
The mean value, $\langle \hat{R}\rangle=1.0008$, indicates consistent sampling across chains for the adopted parametrisation.


The best-fit \texttt{H0\_fixed\_SPEMD} model accurately reproduces the observed multiple-image positions, with a chi-square per degree-of-freedom of $\tilde{\chi}^2_\mathrm{min}=0.64$, and an image-plane rms separation of $\Delta_\mathrm{rms}=0.15''$ between the observed and model-predicted positions. 
A summary of the model performance is presented in the top row of Table~\ref{tab:bestmodel}, while the corresponding marginalised posterior distributions of the model parameters are reported in Table~\ref{tab:modelparams}. 

\begingroup
\setlength{\tabcolsep}{3.8pt}       
\renewcommand{\arraystretch}{1.4}
    \begin{table*}
    \centering
    \small
    \caption{Summary of the best-fit \textsc{Gravity.jl} SL models for M0138.}
    \label{tab:bestmodel}
    \begin{tabular}{l c c c c c c c c c c c c c} \hline \hline
    Model & $N_{\rm par}$ & $N_{\rm const}$ & d.o.f. & $\chi^2_{\mathrm{min, \,tot}}$ & $\chi_{\mathrm{min, \,pos}}^{2}$ & $\chi_{\mathrm{min,\,td}}^{2}$ & $\chi_{\mathrm{min,\,td\,(E)}}^{2}$ & $\chi_{\mathrm{min,\,td\,(R)}}^{2}$ & $\tilde\chi^2_{\mathrm{min}}$ & $\log Z$ & BIC & $\Delta_\mathrm{rms} \, ['']$ & $H_0\,[\mathrm{km}\,\mathrm{s}^{-1}\,\mathrm{Mpc}^{-1}]$ \\
    \hline
    \texttt{H0\_fixed\_SPEMD}  & 21 & 46 &  9 &  5.72 &  5.72 & /    &    / &   /  & 0.64 & $-$208.22 & 147.4 & 0.15 & 70 (fixed) \\ \hline
    \texttt{H0\_free\_SPEMD\_E}& 22 & 47 &  9 &  6.56 &  6.14 & 0.42 & 0.42 &   /  & 0.73 & $-$198.04 & 152.0 & 0.13 & $71.5^{+10.2}_{-8.4}$ \\ \hline
    \texttt{H0\_free\_SPEMD}   & 22 & 49 & 11 & 12.59 &  7.30 & 5.29 & 0.02 & 5.27 & 1.14 & $-$195.17 & 158.1 & 0.12 & $70.2^{+9.5}_{-7.9}$  \\
    \texttt{H0\_free\_NFW}     & 21 & 49 & 12 & 14.89 &  8.21 & 6.68 & 1.35 & 5.33 & 1.24 & $-$196.54 & 156.5 & 0.15 & $65.4^{+8.8}_{-7.3}$ \\
    \texttt{H0\_free\_NIE}     & 21 & 49 & 12 & 16.76 & 10.79 & 5.97 & 0.02 & 5.95 & 1.40 & $-$199.75 & 158.4 & 0.24 & $66.9^{+9.3}_{-7.8}$ \\
    \hline
    \end{tabular}
    \tablefoot{Summary of the SL models and their overall performance. The columns report the model name, the number of free parameters explicitly varied in the lens model, $N_{\rm par}$, the number of observational constraints, $N_{\rm const}$, the number of degrees of freedom, d.o.f., the minimum total chi-square value, $\chi^2_{\mathrm{min, \,tot}} = \chi_{\mathrm{min, \,pos}}^{2} + \chi_{\mathrm{min, \,td}}^{2}$, the positional contribution to the total chi-square, $\chi_{\mathrm{min, \,pos}}^{2}$, the time-delay contribution to the total chi-square, $\chi_{\mathrm{min, \,td}}^{2} = \chi_{\mathrm{min, \,td\,(E)}}^{2} + \chi_{\mathrm{min, \,td\,(R)}}^{2}$ (where the two terms denote the contributions from SN~Encore and SN~Requiem, respectively), the minimum chi-square value per degree-of-freedom $\tilde\chi^2_{\mathrm{min}}$, the logarithm of the Bayesian evidence, $\log Z$, the BIC value, the $\Delta_{\mathrm{rms}}$ value, and the inferred value of $H_0$ (in the \texttt{H0\_fixed\_SPEMD} model $H_0$ instead was fixed to that value). The number of degrees of freedom is computed as ${\rm d.o.f.}=N_{\rm const}-N_{\rm par}-2N_{\rm src}$, where $N_{\rm src}=8$ is the number of independent source families, each contributing two source-plane coordinates that are marginalised over as nuisance parameters.}
    \end{table*}
\endgroup

\begin{figure}
    \centering
     \includegraphics[width=0.99\linewidth]{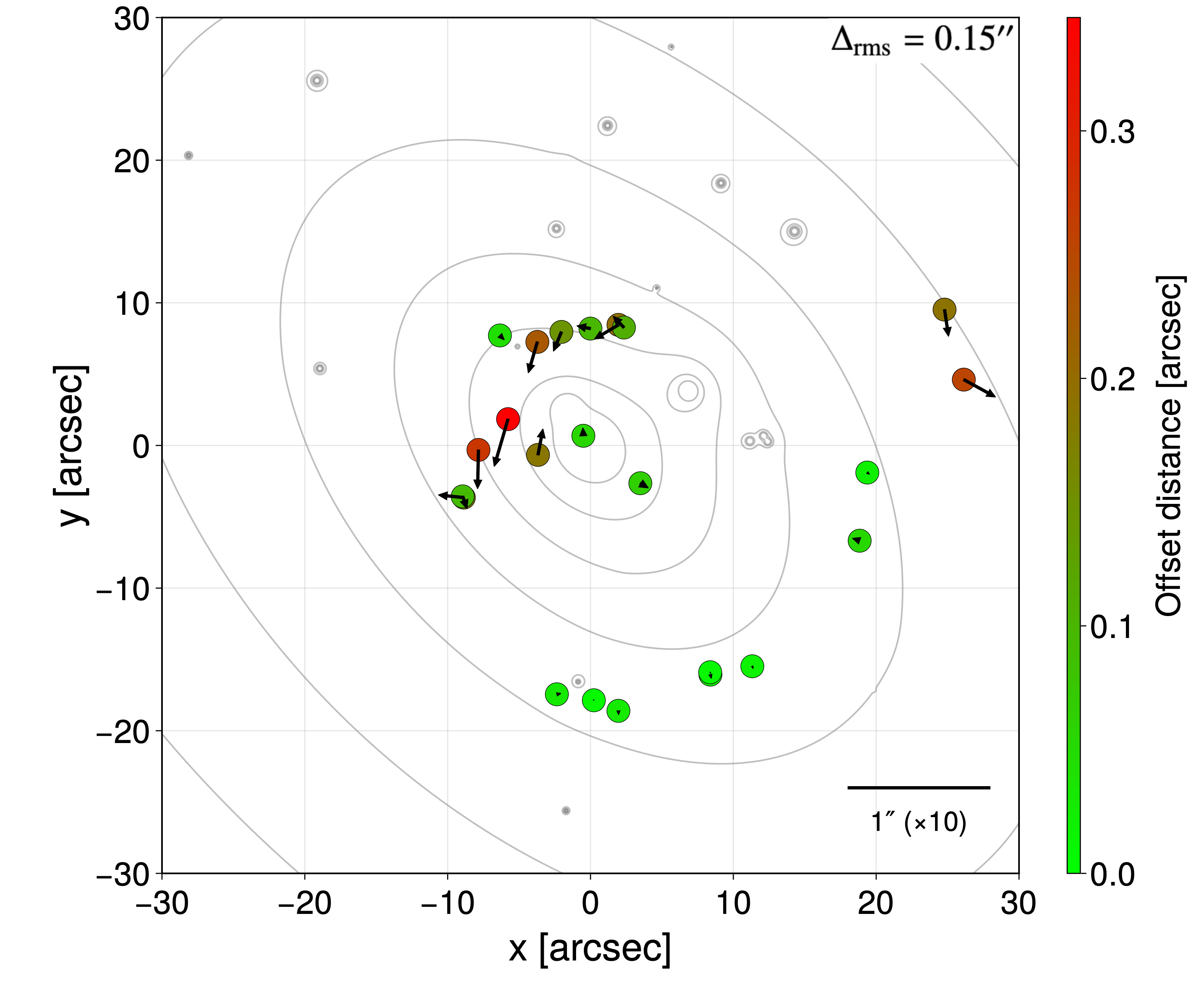}
     \caption{Positional offsets between the observed and model-predicted multiple images in MACS J0138. The circles mark the observed image positions, while the arrows indicate the direction and magnitude of the offsets toward the corresponding model-predicted positions. For visual clarity, the arrow lengths have been magnified by a factor of ten relative to the true positional offsets. The $(0,0)$ point corresponds to the position of the BCG. The background contours show the total projected mass distribution obtained from the best-fitting lens model.}
      \label{fig:scatter}
\end{figure}

Figure~\ref{fig:scatter} shows the positional offsets between the observed and model-predicted multiple images. 
The circles mark the observed image positions, while the arrows point toward the corresponding model-predicted positions. 
For visualisation purposes, the arrow lengths are scaled by a factor of ten relative to the true residuals. 
The residuals do not show a coherent large-scale pattern, suggesting that the adopted model captures the dominant terms in the projected lensing potential. 
The largest offsets occur for multiple images in regions where the lens mapping is less tightly constrained and more sensitive to local perturbations.

\subsection{Reconstruction of the surface-brightness distribution of MRG-M0138}
\label{subsec:sb}

In order to test the validity of our model, we reconstructed the SB distribution of MRG-M0138, the host galaxy of SN Encore and SN Requiem, using \textsc{Gravity.jl}. 
For this purpose, we adopted the best-fitting reference lens model \texttt{H0\_fixed\_SPEMD}, and optimised the parameters of a single Sérsic profile~\citep{sersic63} describing the intrinsic SB distribution of the background source. 
The Sérsic index was fixed to $n=4$~\citep{devaucouleurs48}, while the remaining source-light parameters were inferred in a Bayesian framework by minimising the difference between the observed and model-predicted SB on the image plane.

We optimised the Sérsic parameters using the pixels associated with the two arcs hosting multiple images 3a, 3b, and 3c, together with the corresponding error map produced by the image processing pipeline~\citep{koekemoer11}.  
The fit used $20.6\times10^3$ JWST pixels in the F200W band, with a pixel scale of $0.060''$, to constrain the best-fitting parametric SB model of the background source. 
We used the F200W point-spread function created by~\citet{ertl25} from multiple stars in the field with the software \texttt{STARRED}~\citep{michalewicz23, millon24}. 
The resulting best-fitting Sérsic profile has a circularised effective radius of $R_\mathrm{e}\simeq 1\,\mathrm{kpc}$, in agreement with the results by~\citealt{newman26} (using our best-fit axis-ratio of $b/a=0.28$). 
Using this optimised source-light model and the fixed best-fitting lens model, we then predicted a posteriori the SB distribution of the other counter-images of the host galaxy, without including them directly in the lens-model optimisation. 
The purpose of this reconstruction was therefore to assess the predictive performance of the lens model, rather than to infer the intrinsic structural properties of the source.

\begin{figure*}
    \centering
     \includegraphics[width=0.99\textwidth]{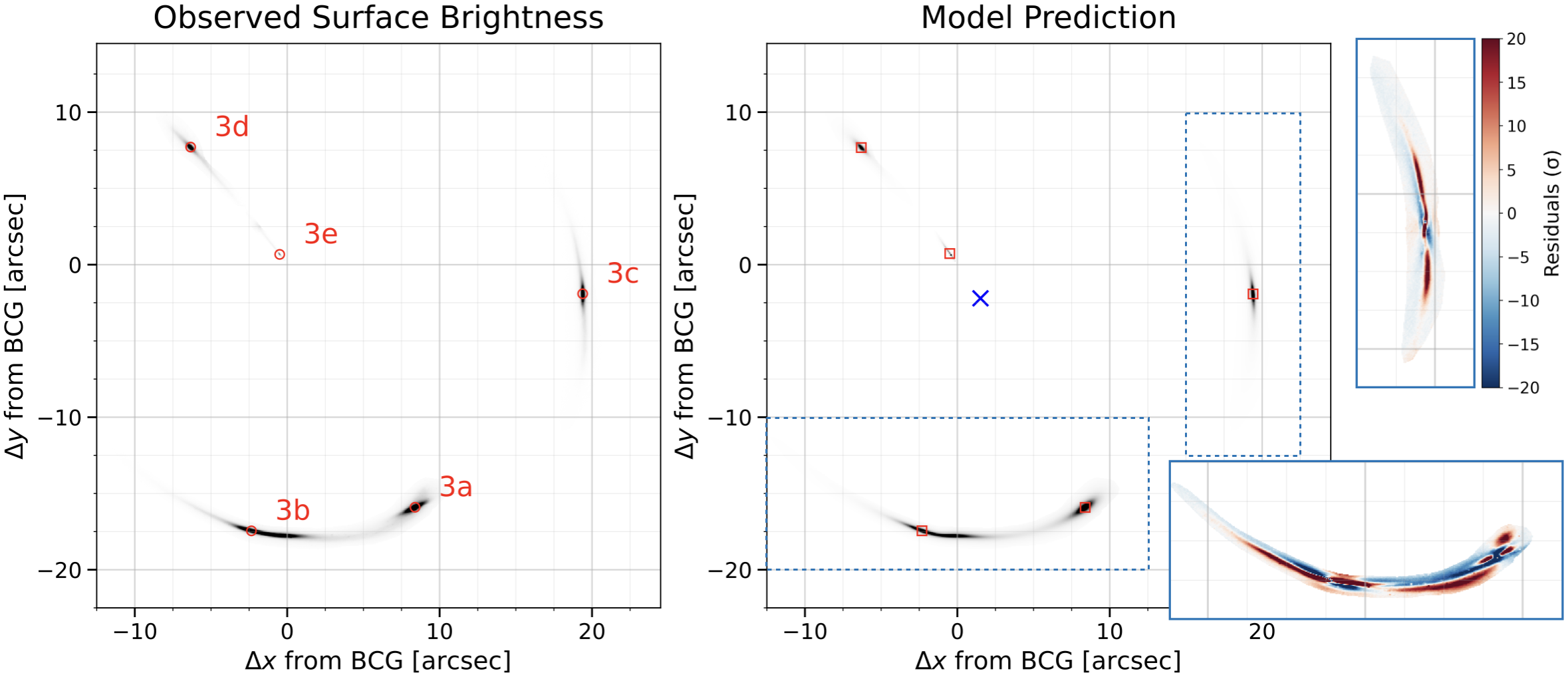}
     \caption{Comparison between the observed and model-predicted SB distribution of the multiply imaged host galaxy MRG-M0138. The left panel shows the observed BCG-subtracted JWST F200W image, while the right panel shows the image-plane prediction obtained by lensing the best-fitting Sérsic model of the background source through the reference \texttt{H0\_fixed\_SPEMD} mass model. The Sérsic source-light parameters were optimised using the pixels associated with the two arcs hosting multiple images 3a, 3b, and 3c. 
     The two insets on the right show the normalised residuals, clipped to the range $[-20\,\sigma,+20\,\sigma]$, for the two regions highlighted by the dashed blue boxes in the model-prediction panel. Residuals associated with host-galaxy images 3d and 3e are omitted because of the contamination from the BCG. Red circles mark the observed multiple images of MRG-M0138 (left panel), listed in Table~4 in~\citet{ertl25}, while the red boxes in the right panel correspond to the predicted positions. The blue cross in the right panel shows the predicted position of the host galaxy in the source plane. Coordinates are given in arcseconds relative to the BCG ($\mathrm{RA} = 24.51570318$ deg, $\mathrm{Dec} = -21.92547911$ deg, given in the International Celestial Reference System frame).}
      \label{fig:model}
\end{figure*}

Figure~\ref{fig:model} compares the observed SB distribution of the multiple images of MRG-M0138 with the corresponding image-plane prediction from our best-fit reference model. 
For visualisation, we used the BCG-subtracted image from~\citet{ertl25}, which improves the visibility of the radial arc of MRG-M0138 and facilitates the comparison with the model-predicted SB. 
The colour-coded normalised residuals shown in Fig.~\ref{fig:model} are compatible with those in~\cite{acebron25} obtained with \textsc{LensTool}~\citep{kneib96, jullo07}. 
The residuals are shown only as a qualitative diagnostic of the SB reconstruction, since the extended SB information is not included in the lens-model optimisation. 
The model reproduces the main morphology and relative SB distribution of the multiple images, providing an additional posterior predictive check of the lens model. 

This comparison is nevertheless limited by the adopted parametric description of the source light and by the fact that the extended SB information is not yet used to optimise the lens mass model. 
A more flexible reconstruction based on a pixelised source grid~\citep[e.g.][]{suyu06, gu22, acebron24, birrer25, schuldt26, huang26a, huang26b} would allow the resolved host-galaxy emission to be incorporated directly into the likelihood, thereby improving the overall lens reconstruction and providing stronger local constraints on the lensing distortion field. 
A full SB optimisation of the lensed host galaxy will be explored in future work.

\subsection{Predicted time delays and magnifications of SN Encore and Requiem}
\label{subsec:predictions}

We used the posterior distribution of the \texttt{H0\_fixed\_SPEMD} model to predict the multiple-image configurations, time delays, and magnifications of SN Encore and SN Requiem. 
For each posterior sample, the observed image positions associated with a given source family were first mapped to the source plane through the lens equation. 
The source position was then estimated as the weighted average of the back-projected source coordinates. 
Starting from this inferred source position, we solved the non-linear lens equation to recover the corresponding model-predicted image positions in the image plane. 
The relative arrival times were obtained by evaluating the Fermat-potential differences at the predicted multiple image positions. 

To derive the predicted position, time-delay, and magnification distributions, we randomly drew $5000$ realisations from the MCMC chains and computed the corresponding observables for each realisation. 
The predicted time delays and magnifications for SN Encore and SN Requiem from the \texttt{H0\_fixed\_SPEMD} model are listed in Table~\ref{tab:SNtimemag}. 
All sampled models predict at least four multiple images for both SNe. 
For SN Encore, the model predicts the two observed images 1a and 1b, a third fainter image 1c, and a highly delayed image 1d. 
For SN Requiem, the model predicts the three observed images 2a, 2b, and 2c, together with a delayed image 2d. 
In a subset of posterior samples, the model also produces an additional central image near the BCG for one or both SNe. 
These central images, labelled 1e and 2e in Table~\ref{tab:SNtimemag}, are not recovered in all the posterior realisations (as in~\citealt{suyu26}) and are expected to be highly demagnified; their predicted properties should therefore be interpreted with caution.

\begin{table}
    \centering
    \caption{Predicted time delays and magnifications of multiple images of SN Encore and SN Requiem (\texttt{H0\_fixed\_SPEMD} mass model).}
    \begin{tabular}{llc}
        \toprule
        Image & Time delay [days] & Magnification $(\mu)$\\ 
        \midrule
        \midrule
        \multicolumn{3}{l}{SN Encore predictions}\\
        \midrule       
         1a              & /                                            & $-24.1^{+2.1}_{-2.2}$ \vspace{2px} \\
         1b              & $\Delta t_{\rm 1b,1a} = -39.4^{+3.0}_{-3.4}$ & $36.2^{+5.4}_{-5.7}$  \vspace{2px} \\
         1c*             & $\Delta t_{\rm 1c,1a} = -376^{+23}_{-24}$    & $10.1^{+0.6}_{-0.6}$  \vspace{2px} \\
         1d*             & $\Delta t_{\rm 1d,1a} = 3122^{+75}_{-70}$    & $-4.0^{+0.4}_{-0.4}$  \vspace{2px} \\
         1e*$^{\dagger}$ & $\Delta t_{\rm 1e,1a} = 3413^{+60}_{-61}$    & $1.58^{+1.8}_{-0.6}$ \\
        \midrule
        \multicolumn{3}{l}{SN Requiem predictions}\\
        \midrule
        2a              & /                                              & $-29.0^{+2.8}_{-3.0}$ \vspace{2px} \\
        2b              & $\Delta t_{\rm 2b,2a} = -57.2^{+5.2}_{-4.5}$   & $22.1^{+1.9}_{-1.9}$  \vspace{2px} \\
        2c              & $\Delta t_{\rm 2c,2a} = -96.3.0^{+5.9}_{-6.4}$ & $14.1^{+1.3}_{-1.4}$  \vspace{2px} \\
        2d*             & $\Delta t_{\rm 2d,2a} = 3902^{+72}_{-67}$      & $-3.6^{+0.6}_{-0.6}$  \vspace{2px} \\
        2e*$^{\dagger}$ & $\Delta t_{\rm 2e,2a} = 3977^{+71}_{-63}$      & $1.7^{+0.4}_{-1.0}$  \\
        \bottomrule
    \end{tabular}
    \label{tab:SNtimemag}
    \tablefoot{Values are reported as the median from the $5000$ sampled chains, with uncertainties corresponding to the $16$th and $84$th percentiles. Images marked with * are model-predicted images that have not been detected to date, either because of their faintness, as for image 1c, or because of their long predicted time delays, as for images 1d, 1e, and 2d. Image 1e of SN Encore and Image 2e of SN Requiem, marked with $^{\dagger}$, are predicted in $73\%$ and $18\%$ of the posterior samples, respectively; all other images are predicted in all samples. The reference anchor epoch $t=0$ is set to 2016-07-18 for image 2a of SN Requiem~\citep{rodney21}, and to 2023-11-17 for image 1a of SN Encore~\citep{pierel24}.}
\end{table}

For SN Encore, the model predicts a time delay $\Delta t_{\rm 1b,1a}=-39.4^{+3.0}_{-3.1}$ days, in agreement with the measured value of $-39.8^{+3.9}_{-3.3}$ days from the SN light-curve analysis~\citep{pierel26}. 
Image 1d is predicted to appear after image 1a with a time delay of $\Delta t_{\rm 1d,1a}=3122^{+75}_{-70}$ days and a magnification of $\mu_{\rm 1d}=-4.0^{+0.4}_{-0.4}$. 

\begin{figure}
    \centering
    \includegraphics[width=0.99\linewidth]{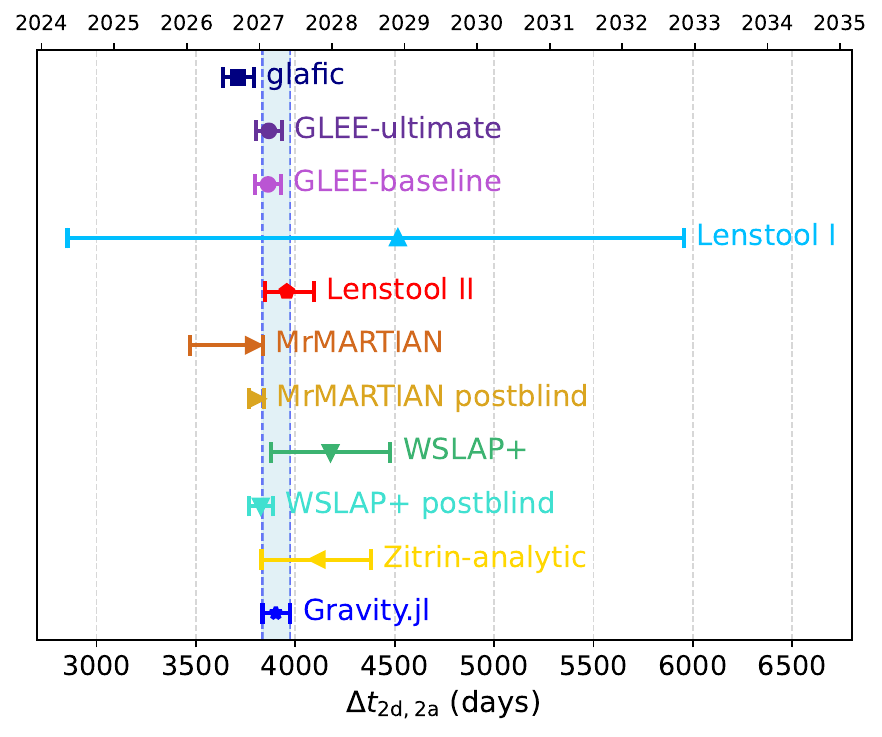}
    \caption{Predicted reappearance time of image 2d of SN Requiem, expressed as the time delay relative to image 2a, $\Delta t_{\rm 2d,2a}$. The blue point shows the prediction from the \textsc{Gravity.jl} model presented in this work, while the other points show the predictions from the independent lens models analysed by~\citet{suyu26}. Horizontal error bars indicate the corresponding $68\%$ credible intervals. For our model, the interval is derived from $5000$ posterior samples randomly drawn from the \texttt{H0\_fixed\_SPEMD} MCMC chains.}
    \label{fig:dt2d}
\end{figure}

For SN Requiem, the delayed image 2d is predicted to appear after image 2a with a delay of $\Delta t_{\rm 2d,2a}=3902^{+72}_{-67}$ days and a magnification $\mu_{\rm 2d}=-3.6^{+0.6}_{-0.6}$. 
Figure~\ref{fig:dt2d} compares this predicted reappearance time with the independent lens models analysed by~\citet{suyu26}. 
Our \textsc{Gravity.jl} prediction is consistent with the estimates presented in that work within the current modelling uncertainties, providing an independent cross-check of the Fermat-potential reconstruction in the region of the SN Requiem images.

\begin{figure*}
    \centering
     \includegraphics[width=0.99\linewidth]{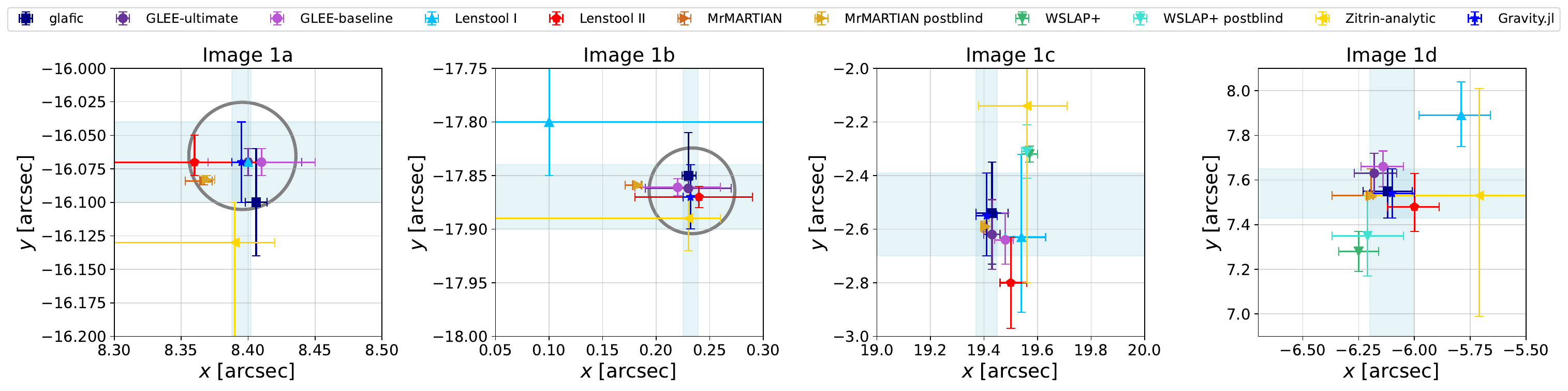}
     \caption{Comparison of the predicted positions of the multiple images of SN Encore from the \textsc{Gravity.jl} model with those obtained from the independent lens models analysed by~\citet{suyu26}. Each panel shows one image of SN Encore: the observed images 1a and 1b, and the model-predicted images 1c and 1d. Different symbols and colours correspond to different lens models, as indicated in the legend. The \textsc{Gravity.jl} predictions are shown in blue. The shaded bands show the projected uncertainty region associated with the \textsc{Gravity.jl} prediction. Grey circles denote the observed measured values~\citep{pierel24}.}
      \label{fig:pos_encore}
\end{figure*}

\begin{figure*}
    \centering
    \includegraphics[width=0.99\linewidth]{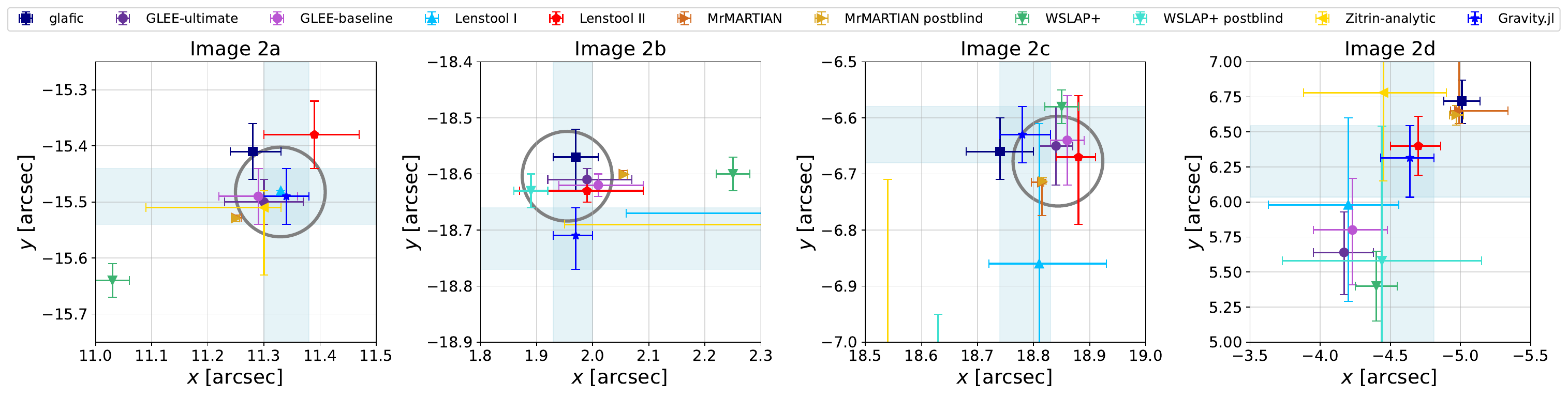}
    \caption{Comparison of the predicted positions of the multiple images of SN Requiem from the \textsc{Gravity.jl} model with those obtained from the independent lens models analysed by~\citet{suyu26}. Each panel shows one image of SN Requiem: the observed images 2a, 2b, and 2c, and the delayed image 2d. Different symbols and colours correspond to different lens models, as indicated in the legend. The \textsc{Gravity.jl} predictions are shown in blue. The shaded bands show the projected uncertainty region associated with the \textsc{Gravity.jl} prediction. Grey circles denote the observed measured values~\citep{rodney21}.}
    \label{fig:pos_requiem}
\end{figure*}

Figures~\ref{fig:pos_encore} and~\ref{fig:pos_requiem} compare the model-predicted positions of the multiple images of SN Encore and SN Requiem with those obtained from the independent lens models analysed by~\citet{suyu26}. 
The \textsc{Gravity.jl} predictions are consistent with the observed positions of SN Encore images 1a and 1b, and of SN Requiem images 2a, 2b, and 2c. 
They also fall within the range of image-position predictions obtained from the independent lens models analysed by~\citet{suyu26}, supporting the robustness of the predicted SN Requiem reappearance configuration within the current modelling uncertainties. 

Similarly, the predicted magnifications for the multiple images of both SNe are in good agreement with those reported by~\citet{suyu26}, within the quoted uncertainties. 
As a consistency check, we used the predicted magnification of image 2a of SN~Requiem, $\mu_{\rm 2a} = -29.0^{+2.8}_{-3.0}$ (cf.~Table~\ref{tab:SNtimemag}), together with its observed HST F160W magnitude close to the light curve peak, $m_\mathrm{F160W} = 22.52 \pm 0.01 $~\citep{rodney21}. 
After correcting for lensing magnification, we inferred an absolute magnitude at rest-frame $\simeq0.5\,\mu\mathrm{m}$ of $M = m_\mathrm{F160W} + 2.5 \log_{10}|\mu_{\rm 2a}| - \mathrm{DM} = -19.7^{+0.1}_{-0.1}$, where $\mathrm{DM}=45.9$ is the distance modulus at the SN redshift. 
This value is well consistent with the typical rest-frame optical peak luminosities of normal Type~Ia SNe~\citep[e.g.][]{taubenberger17}. 
This agreement is noteworthy because the lower magnification used by~\citet{rodney21}, based on a preliminary lens model, led to a discrepancy with the typical Type~Ia SN luminosity exceeding one magnitude. 

The residual differences among the predictions shown in Figs.~\ref{fig:dt2d}, \ref{fig:pos_encore}, and~\ref{fig:pos_requiem} should be interpreted as an estimate of the model-dependent systematic uncertainties affecting the predicted reappearance positions and time delays, complementary to the statistical uncertainties inferred from the posterior distribution of any individual lens model. 

\subsection{Measurement of $H_0$ with SNe Requiem and Encore}
\label{subsec:free}

We next inferred $H_0$ by including the measured time delays of SN~Encore and SN~Requiem, together with their uncertainties, in the SL likelihood. 
Specifically, for the \texttt{H0\_free\_SPEMD\_E} model, we included only the SN~Encore time delay, $\Delta t_{\rm 1b,1a}$, as a constraint, whereas for \texttt{H0\_free\_SPEMD}, our reference cosmographic model, we also included the two SN~Requiem time delays, $\Delta t_{\rm 2a,2b}$ and $\Delta t_{\rm 2c,2b}$ (cf.~Sect.~\ref{sec:modeling}).

In TDC, the relative arrival time between two multiple images depends on both the difference in Fermat potential ($\tau$), and the time-delay distance ($D_{\Delta t}$), 
\begin{equation}
    \Delta t_{ij} =
    \frac{D_{\Delta t}}{c}
    \left[
    \tau(\boldsymbol{\vartheta}_i,\boldsymbol{\beta}) -
    \tau(\boldsymbol{\vartheta}_j,\boldsymbol{\beta})
    \right] \,,
\end{equation}
where $\boldsymbol{\vartheta}_i$ and $\boldsymbol{\vartheta}_j$ are the image positions, and $\boldsymbol{\beta}$ is the source position~\citep[e.g.][]{treu16, birrer24}. 
For a single lens plane, the time-delay distance is
\begin{equation}
    D_{\Delta t} = (1+z_{\mathrm l})\frac{D_{\mathrm l}\,D_{\mathrm s}}{D_{\mathrm{ls}}} \,,
\end{equation}
which scales to leading order as $H_0^{-1}$, for fixed $\Omega_{\mathrm m}$ and $\Omega_\Lambda$. 
The measured SN time delays therefore constrain $H_0$ through the combination of the cosmological distance scale and the lens-model prediction for the Fermat-potential differences~\citep{refsdal64}.

For both cosmographic models, we allowed $H_0$ to vary while keeping the remaining cosmological parameters fixed to the fiducial values adopted throughout this work. 
We retained the same image-position constraints, total mass parametrisation, and priors adopted for the fixed-cosmology \texttt{H0\_fixed\_SPEMD} model, with the addition of a uniform prior $H_0 \sim \mathcal{U}(20,120)\,\mathrm{km}\,\mathrm{s}^{-1}\,\mathrm{Mpc}^{-1}$ (cf.~Table~\ref{tab:modelparams}).

The best-fitting \texttt{H0\_free\_SPEMD} model has a reduced chi-square of $\tilde{\chi}^2_\mathrm{min}=1.14$, an image-plane positional rms of $\Delta_\mathrm{rms}=0.12''$, and a mean chain convergence statistic of $\langle\hat{R}\rangle=1.0004$, as reported in Table~\ref{tab:bestmodel}, together with the results for the \texttt{H0\_free\_SPEMD\_E} model. 
The increase in $\chi^2_\mathrm{min,\,tot}$ for \texttt{H0\_free\_SPEMD} relative to \texttt{H0\_fixed\_SPEMD} is driven primarily by the time-delay contribution to the likelihood rather than by a degradation of the image-position fit. 
In particular, the $\chi^2_{\rm min,td}$ term is dominated by the SN~Requiem time-delay contribution (cf.~Table~\ref{tab:bestmodel}). 
Nevertheless, the values of $H_0$ inferred from the \texttt{H0\_free\_SPEMD} and \texttt{H0\_free\_SPEMD\_E} models are fully consistent within $1\,\sigma$, since the cosmographic inference is driven primarily by the more precise SN~Encore time-delay measurement. 
The inferred mass parameters also remain consistent with those of \texttt{H0\_fixed\_SPEMD} (cf.~Table~\ref{tab:modelparams}), indicating that the time-delay constraints primarily calibrate the overall cosmological distance scale rather than requiring a substantial change in the reconstructed total projected mass distribution. 
The marginalised posterior distributions of the main lens-model parameters and $H_0$ for the \texttt{H0\_free\_SPEMD} model are shown in Fig.~\ref{fig:corner} in Appendix~\ref{apdx:C}.

From the joint analysis of SN~Encore and SN~Requiem, we infer $H_{0, \mathrm{SPEMD}} = 70.2^{+9.5}_{-7.9}\,\mathrm{km}\,\mathrm{s}^{-1}\,\mathrm{Mpc}^{-1}$, while the Encore-only model gives $H_{0, \mathrm{SPEMD\_E}}=71.5^{+10.2}_{-8.4}\,\mathrm{km}\,\mathrm{s}^{-1}\,\mathrm{Mpc}^{-1}$. 
Adding the measured time delays between the SN~Requiem multiple images therefore produces a small shift of the posterior toward lower values of $H_0$, but does not significantly reduce the width of the distribution. 
This behaviour is expected because the SN~Encore delay is measured more precisely, with $\Delta t/t\simeq10\%$, whereas the SN~Requiem delays are less constraining, with $\Delta t/t\simeq20\%$. 
The inclusion of the Requiem delays nevertheless provides an additional, independent consistency check on the reconstructed Fermat-potential differences.


To obtain a first estimate of the sensitivity of the inferred $H_0$ to the adopted mass parametrisation, we repeated the MCMC analysis for two variants of the reference \texttt{H0\_free\_SPEMD} model. 
These models use the same observational constraints and mass parametrisation as the reference model, but adopt different parametrisations for the large-scale cluster halo. 
Specifically, we considered 
\begin{itemize}
    \item the \texttt{H0\_free\_NFW} model, in which the cluster halo is described by an elliptical NFW profile, with six free mass parameters;
    \item the \texttt{H0\_free\_NIE} model, in which the cluster halo is described by a NIE profile, also with six free mass parameters.
\end{itemize}
For both models, we adopted the same sampler configuration as for the reference \texttt{H0\_free\_SPEMD} model. 
We obtained $H_{0, \mathrm{NFW}} = 65.4^{+8.7}_{-7.3}\,\mathrm{km}\,\mathrm{s}^{-1}\,\mathrm{Mpc}^{-1}$
for the NFW model, and $H_{0, \mathrm{NIE}} = 66.9^{+9.3}_{-7.8}\,\mathrm{km}\,\mathrm{s}^{-1}\,\mathrm{Mpc}^{-1}$ for the NIE model (cf.~Table~\ref{tab:bestmodel}). 
The fractional uncertainty on $H_0$ is comparable for all three halo parametrisations, and the shifts in the median inferred values remain well within the statistical uncertainties of the individual posteriors. 
This suggests that, for the present dataset and within the limited set of mass parametrisations explored here, the $H_0$ inference remains dominated by the statistical uncertainty of the measured SN time delays rather than by the model-to-model scatter. 
These tests should therefore be regarded as a first robustness assessment rather than as a complete exploration of lens-model systematics. 
A complete assessment of model-dependent systematics is, however, beyond the scope of the present work and will be addressed in a future analysis, when additional data from the ongoing monitoring campaign become available. 

To propagate the uncertainty associated with the choice of mass-model parametrisation into the final $H_0$ inference, we combined the three posterior distributions using Bayesian model averaging~\citep[e.g.,][for a review]{hoeting99,parkinson13}. 
For each lens model $M_i$, we assigned the posterior model probability $P(M_i\mid D)$ as
\begin{equation}
    w_i \equiv P(M_i\mid D) = \frac{Z_i P(M_i)}{\sum_j Z_j P(M_j)}\,,
\end{equation}
where $Z_i=P(D\mid M_i)$ is the Bayesian evidence, $P(M_i)$ is the prior probability assigned to model $M_i$, and $D$ denotes the data~\citep{ivezic20}. 
Assuming equal prior probabilities for the three models, the weights reduce to $w_i=Z_i/\sum_j Z_j$. 
For numerical stability, since the evidence values can be extremely small, we evaluated the weights using the log-evidences
\begin{equation}
    w_i = \frac{\exp(\log Z_i-\log Z_{\max})}{\sum_j \exp(\log Z_j-\log Z_{\max}) } \,.
\end{equation} 

The model-averaged $H_0$ posterior was then constructed as the finite mixture 
\begin{equation}
    p(H_0\mid D) = \sum_{i=1}^{3} w_i\,p(H_0\mid D,M_i) \,.
\end{equation}
For posterior distributions represented by $N_i$ equally weighted samples, each sample from model $M_i$ was assigned a weight $w_i/N_i$. 
This normalisation ensures that the integrated contribution of each model is equal to $w_i$, independently of the number of samples in its posterior chain, and that the weights of all samples across the three models sum to unity. 
The resulting model-averaged posterior of $H_0$ therefore accounts for both the statistical uncertainty within each individual model and the model-to-model uncertainty associated with the three mass parametrisations considered here, encoded through the Bayesian evidence weights.

The Bayesian model averaging of the three posterior distributions yields a final constraint on the value of the Hubble constant of
\begin{equation}
    H_0 = 69.2^{+9.6}_{-8.0}\,\mathrm{km}\,\mathrm{s}^{-1}\,\mathrm{Mpc}^{-1}\,.
\end{equation}
This value of $H_0$ is consistent within $1\,\sigma$ with the constraint obtained by~\citet{suyu26, pierel26}, who combined seven independent SL models of M0138 and inferred $H_0=66.9^{+11.2}_{-8.1}\,{\rm km\,s^{-1}\,Mpc^{-1}}$. 
In contrast to the analysis of~\citet{suyu26}, where the cosmographic inference was performed a posteriori from fixed mass-model realisations, here $H_0$ is sampled jointly with the lens-model parameters, and the two additional measured time delays of SN~Requiem were included as constraints. 

\begin{figure}
    \centering
    \includegraphics[width=0.99\linewidth]{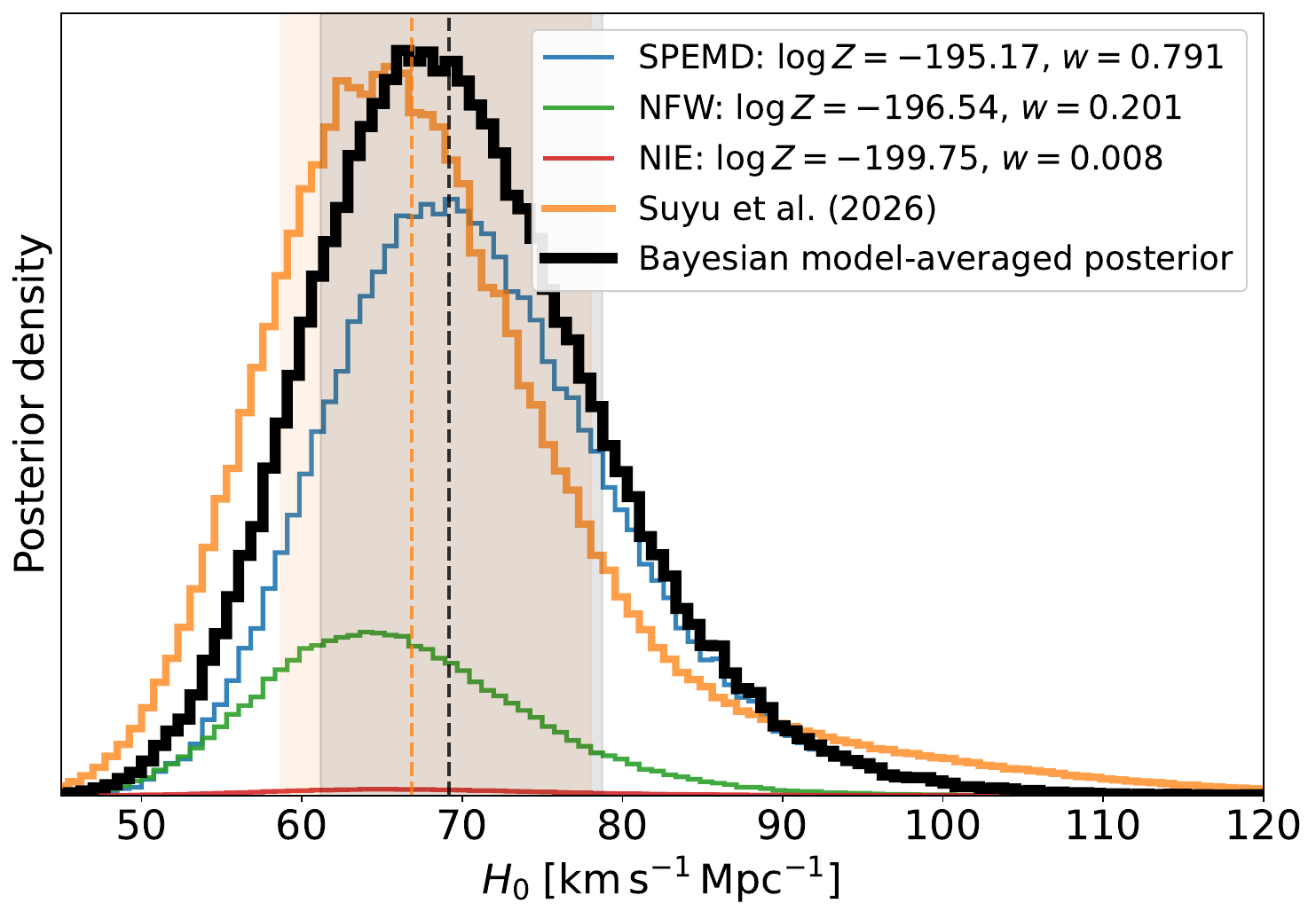}   
    \caption{Marginalised posterior distributions of $H_0$ inferred for M0138. The blue, green, and red distributions show the constraints obtained with the \texttt{H0\_free\_SPEMD}, \texttt{H0\_free\_NFW}, and \texttt{H0\_free\_NIE} models, respectively. The black distribution shows the Bayesian model-averaged posterior obtained by combining these three \textsc{Gravity.jl} lens models using their evidence-based model probabilities. For comparison, the orange distribution shows the $H_0$ constraint inferred by~\citet{suyu26, pierel26} from a weighted combination of seven independent SL models of M0138. Vertical lines mark the median values of the corresponding posterior distributions, while shaded regions show credible intervals.}
    \label{fig:H0_posteriors_comparison}
\end{figure}

Figure~\ref{fig:H0_posteriors_comparison} compares the $H_0$ posteriors obtained with the \texttt{H0\_free\_SPEMD}, \texttt{H0\_free\_NFW}, and \texttt{H0\_free\_NIE} models, together with their Bayesian model-averaged posterior. 
For comparison, we also show the posterior inferred by~\citet{suyu26, pierel26}. 
All posteriors are mutually consistent within their quoted uncertainties. 
The comparison further indicates that, at the present level of precision, the $H_0$ error budget is dominated by the statistical uncertainty of the measured SN time delays, $\Delta t/t\gtrsim10\%$, rather than by the scatter among the three cluster-scale halo parametrisations explored here. 
This conclusion is consistent with previous systematic tests of M0138~\citep{ertl25, acebron25}; in particular, \citet{acebron25} found that the model-to-model scatter in the predicted SN time delays remained within the statistical uncertainties of the individual predictions for the mass models considered in that work.




\section{Summary and conclusions}
\label{sec:conclusions}


We have presented a SL analysis of the galaxy cluster MACS~J0138.0$-$2155, which lenses the massive red galaxy MRG-M0138 at $z_\mathrm{s}=1.949$, the host of the two multiply imaged Type Ia SNe Requiem and Encore. 
This system provides a rare opportunity to test cluster lens models with multiple lensed transients in the same background galaxy and to infer the value of the Hubble constant from independent SN time-delay measurements. 
Our analysis was carried out with the \textsc{Gravity.jl} lens-modelling framework~\citep{lombardi24} and uses the spectroscopically confirmed ``gold'' sample of multiple-image systems identified from HST, JWST, and MUSE data~\citep{ertl25}, together with cluster-member scaling relations calibrated with stellar kinematics~\citep{granata25}.

Our reference fixed-cosmology model, \texttt{H0\_fixed\_SPEMD}, accurately reproduces the observed multiple-image positions, yielding a reduced chi-square of $\tilde{\chi}^2_{\rm min}=0.64$ and an image-plane rms offset of $\Delta_{\rm rms}=0.15''$, which is currently the smallest value obtained when compared with all other models presented in~\citet{suyu26}. 
The residuals (cf.~Fig.~\ref{fig:scatter}) do not show a coherent large-scale pattern, indicating that the adopted parametrisation captures the dominant components of the projected lensing potential. 
As an additional posterior predictive check, we reconstructed the SB distribution of MRG-M0138 using a single Sérsic profile for the intrinsic source light and the best-fit reference mass model. 
The resulting image-plane prediction reproduces the overall morphology, position, and relative SB distribution of the observed counter-images, although the extended SB information has not yet been included directly in the lens-model optimisation.

Using posterior samples from the \texttt{H0\_fixed\_SPEMD} model, we predicted the image configurations, time delays, and magnifications of SN~Encore and SN~Requiem. 
For SN~Encore, the model predicts $\Delta t_{\rm 1b,1a}=-39.4^{+3.0}_{-3.4}$ days, in agreement with the measured value of $-39.8^{+3.9}_{-3.3}$ days from the SN light-curve analysis~\citep{pierel26}. 
The model also predicts a highly delayed image 1d, with $\Delta t_{\rm 1d,1a}=3122^{+75}_{-70}$ days and $\mu_{\rm 1d}=-4.0^{+0.4}_{-0.4}$. 
For SN~Requiem, the delayed image 2d is predicted to appear after image 2a with $\Delta t_{\rm 2d,2a}=3902^{+72}_{-67}$ days and $\mu_{\rm 2d}=-3.6^{+0.6}_{-0.6}$. 
The magnification inferred for image 2a, which is estimated to be observed close to the light-curve peak~\citep{rodney21}, leads to a rest-frame optical peak luminosity in very good agreement with the normal Type~Ia SN population~\citep{taubenberger17}, improving on the luminosity estimate by~\citet{rodney21}. 
The predicted reappearance times and image positions are consistent, within the current modelling uncertainties, with the independent lens models analysed by~\citet{suyu26}. 
This agreement supports the robustness of the inferred reappearance configurations of SN~Requiem and SN~Encore within the present modelling uncertainties.


We then inferred the value of $H_0$ by including the measured time delays of SN~Encore and SN~Requiem directly in the SL likelihood. 
In these models, the Hubble constant was allowed to vary while $\Omega_{\mathrm m}$ and $\Omega_\Lambda$ were kept fixed to the fiducial values adopted in this work. 
Our $H_0$ constraint is therefore conditional on the adopted flat $\Lambda$CDM cosmology; marginalising over the remaining cosmological parameters would require additional independent constraints or a larger number of multiply imaged sources spanning a broader range of source redshifts~\citep[e.g.,][]{caminha22}.
The inclusion of the time-delay constraints preserves the quality of the image-plane reconstruction, yielding a positional rms comparable to that of the fixed-cosmology model.  
For the SPEMD parametrisation, the joint SN~Encore and SN~Requiem analysis gives $H_{0, \mathrm{SPEMD}} = 70.2^{+9.5}_{-7.9}\,\mathrm{km}\,\mathrm{s}^{-1}\,\mathrm{Mpc}^{-1}$  (cf.~Table~\ref{tab:modelparams}). 
To obtain a first estimate of the sensitivity of the cosmographic inference to the adopted cluster-scale mass parametrisation, we repeated the analysis using two alternative descriptions of the large-scale halo: an elliptical NFW profile, and a non-singular isothermal ellipsoid. 
These models yield $H_{0, \mathrm{NFW}} = 65.4^{+8.7}_{-7.3}\,\mathrm{km}\,\mathrm{s}^{-1}\,\mathrm{Mpc}^{-1}$, and $H_{0, \mathrm{NIE}} = 66.9^{+9.3}_{-7.8}\,\mathrm{km}\,\mathrm{s}^{-1}\,\mathrm{Mpc}^{-1}$, respectively. 
The fractional uncertainty on $H_0$ is comparable for all three mass parametrisations, and the shifts in their median values remain well within the statistical uncertainties of the individual posteriors. 
This indicates that, for the present dataset and within the family of cluster-scale halo profiles explored here, the $H_0$ inference remains dominated by the statistical uncertainty of the measured SN time delays rather than by the model-to-model scatter.

The Bayesian evidence and $\Delta_{\rm rms}$ values both favour the SPEMD model over the NFW and NIE parametrisations (cf. Table~\ref{tab:modelparams}). 
Rather than relying on this single preferred model, we can combine the three $H_0$ posterior distributions using Bayesian model averaging, weighting each lens model by its Bayesian evidence. 
This procedure propagates both the uncertainty within each individual lens model and the uncertainty associated with the choice among the three cluster-scale halo parametrisations considered here. 
The resulting model-averaged constraint is
\begin{equation}
    H_0 = 69.2^{+9.6}_{-8.0}\,
    \mathrm{km}\,\mathrm{s}^{-1}\,\mathrm{Mpc}^{-1} \, .
\end{equation}
This value is consistent within $1\,\sigma$ with the constraint inferred by~\citet{suyu26, pierel26} from a combination of seven independent SL models of M0138. 
The agreement between the individual \textsc{Gravity.jl} models, their model-averaged posterior, and the independent analyses further indicates that the current error budget is dominated by the relatively large fractional uncertainties of the available SN time delays, $\Delta t/t\gtrsim10\%$. 

The present $H_0$ measurement should nevertheless be interpreted within the set of lens-model families and cosmological assumptions explored here. 
While the Bayesian model averaging accounts for the uncertainty associated with the three alternative cluster-scale halo parametrisations, it does not constitute a complete marginalisation over all lens-model systematics. 
Additional sources of uncertainty include the treatment of galaxy-scale substructure, the contribution of line-of-sight perturbers and external convergence, the adopted scaling relations for cluster members, and residual lensing degeneracies such as the mass-slope and mass-sheet transformations~\citep[e.g.][]{falco85, kawano06, schneider13}. 
A more comprehensive assessment of these systematic effects is beyond the scope of the present work and will be addressed in future analyses~\citep[e.g.][]{grillo20, liu25}. 

Future observations and modelling developments can substantially improve the cosmographic precision of M0138. 
Additional spectroscopy of cluster members and line-of-sight galaxies, together with ongoing efforts to identify and confirm new multiple-image systems, will improve the dynamical and environmental characterisation of the lens and provide additional constraints on the cluster mass distribution. 
In parallel, future work will incorporate the full extended SB distribution of the multiply imaged SN host galaxy directly into the lens-model optimisation, following recent developments in cluster-scale modelling with resolved arcs~\citep[e.g.][]{acebron24, urcelay26, schuldt26}. 
This additional information will provide local constraints on the lensing potential near the SN images, helping to reduce residual degeneracies in the Fermat-potential reconstruction and improve the robustness of the inferred time-delay distance. 

The long predicted delays of the future SN images make M0138 particularly valuable for cluster-scale TDC. 
In particular, the expected reappearance of SN~Requiem image 2d in $\approx 2027$ provides an immediate opportunity for an independent $H_0$ measurement, with a temporal baseline of approximately $11$ years relative to image 2a~\citep{suyu26}. 
This long baseline substantially reduces the fractional impact of the timing uncertainty: even an uncertainty of order one month would correspond to a fractional time-delay uncertainty of only $\approx 1\%$. 
As discussed by~\citet{suyu26}, such a measurement could lead to a statistical precision of $\approx 3$--$5\%$ on $H_0$, making M0138 competitive with other leading cosmological probes, provided that lens-model systematics are sufficiently controlled. 
The forthcoming reappearance of SN~Requiem therefore offers a timely and independent test of the Hubble tension, while the presence of two lensed Type~Ia SNe in the same host galaxy provides additional leverage for validating the cluster lens model. 
M0138 is therefore a benchmark system for cluster-scale TDC and a useful test case for the high-precision SL analyses expected in the JWST, Vera C.~Rubin Observatory~\citep{ivezic19}, Euclid~\citep{mellier25}, and Nancy Grace Roman Space Telescope~\citep{green12} era. 

The MCMC chains and data products generated in this work, including the magnification, convergence, and deflection-angle maps, as well as the best-fitting model-parameter values, will be publicly released upon publication\footnote{\url{https://www.fe.infn.it/astro/lensing/}}. 

\begin{acknowledgements}
We thank the anonymous referee for the helpful and insightful comments that improved the paper. 
LB is indebted to the communities behind the multiple free, libre, and open-source software packages on which we all depend. 
LB thanks Xiaosheng Huang, Massimo Pascale, Luca Pagano, Nicola Barbieri, and Gery A.~Diaz Rubio for insightful discussions on MCMC convergence and Bayesian analysis. 
AA acknowledges financial support through the Beatriz Galindo programme and the project PID2022-138896NB-C51 (MCIU/AEI/MINECO/FEDER, UE), Ministerio de Ciencia, Investigación y Universidades. 
PB acknowledges financial support through grant PRIN-MIUR 2020SKSTHZ and support from the Italian Space Agency (ASI) through contract ``Euclid - Phase E'', INAF Grants ``The Big-Data era of cluster lensing'' and ``Probing Dark Matter and Galaxy Formation in Galaxy Clusters through Strong Gravitational Lensing''. 
ML and CG acknowledge support from the National Recovery and Resilience Plan (NRRP), funded by the European Union – NextGenerationEU; Project title ``GRAVITY'', project code PNRR\_BAC24MLOMB\_01, CUP C53C22000350006. 
MB acknowledges the Department of Physics and Earth Science of the University of Ferrara for the financial support through the FIRD 2025 grant. 
JMD acknowledges the support of projects PID2022-138896NB-C51 (MCIU/AEI/MINECO/FEDER, UE) Ministerio de Ciencia, Investigación y Universidades and SA101P24. 
MJJ acknowledges support for the current research from the National Research Foundation (NRF) of Korea under the program 2022R1A2C1003130. 
This work is based [in part] on observations made with the NASA/ESA/CSA James Webb Space Telescope. The data were obtained from the Mikulski Archive for Space Telescopes at the Space Telescope Science Institute, which is operated by the Association of Universities for Research in Astronomy, Inc., under NASA contract NAS 5-03127 for JWST. These observations are associated with program GO-2345 and DD-6549. Support for US investigators in program GO-2345 and DD-6549 was provided by NASA through a grant from the Space Telescope Science Institute, which is operated by the Association of Universities for Research in Astronomy, Inc., under NASA contract NAS 5-03127.
This research is based on observations made with the NASA/ESA {\it Hubble Space Telescope} obtained from the Space Telescope Science Institute, which is operated by the Association of Universities for Research in Astronomy, Inc., under NASA contract NAS 5–26555. These observations are associated with programs 14496, 15663 and 16264.
The data described here may be obtained from the MAST archive at 
\url{https://doi.org/10.17909/snj9-an10}.
\end{acknowledgements}

\bibliographystyle{aa}
\bibliography{bibliography}

\begin{appendix}
\onecolumn

\section{Model parameters}
\label{apdx:B}

In Table~\ref{tab:modelparams}, we list the posterior values of the model parameters for the two lens models, \texttt{H0\_fixed\_SPEMD} and \texttt{H0\_free\_SPEMD}. 

\begin{table}[ht!]
\centering
\fontsize{9}{9}\selectfont
\caption[]{Model parameter priors and inferred posteriors. }
\begin{tabular}{lllrr} \toprule
Parameter & Description & Prior & \texttt{H0\_fixed\_SPEMD} & \texttt{H0\_free\_SPEMD} \\ \midrule \midrule

\multicolumn{5}{l}{\textbf{Cosmological parameters}} \\
\midrule
$H_0\, [\mathrm{km/s/Mpc}]$ & Hubble constant & fixed/$\mathcal{U}$($20,120$) & $70$ & $70.2_{-7.9}^{+9.5}$
\vspace{2px}\\  

\midrule
\multicolumn{5}{l}{\textbf{Cluster members in scaling relations} (Spherical dPIE; $z=z_{\mathrm{l}}$)} \\
\midrule
$\sigma_{0}\ [\textrm{km/s}]$ & Velocity dispersion & $\mathcal{N}$($330.1,29.0$) & $306_{-23}^{+28}$   & $302_{-22}^{+27}$   \vspace{2px}\\  
$r_\mathrm{cut,0}\ [\arcsec]$ & Truncation radius   & $\mathcal{U}$($1,100$)      & $9.4_{-5.3}^{+8.0}$ & $11.5_{-6.3}^{+9.6}$ \vspace{2px}\\  

\midrule
\multicolumn{5}{l}{\textbf{Dark matter halo} (SPEMD; $z=z_{\mathrm{l}}$)} \\
\midrule
$x_\mathrm{DM}\ [\arcsec]$            & $x$ centroid        & $\mathcal{U}$($-7,7$)     & $-0.1_{-0.2}^{+0.3}$    & $-0.1_{-0.2}^{+0.3}$ \vspace{2px}\\
$y_\mathrm{DM}\ [\arcsec]$            & $y$ centroid        & $\mathcal{U}$($-7,7$)     & $-0.6_{-0.4}^{+0.3}$    & $-0.6_{-0.4}^{+0.3}$ \vspace{2px}\\
$q_\mathrm{DM}$                       & Axis ratio          & $\mathcal{U}$($0.1,1$)    & $0.59_{-0.03}^{+0.03}$  & $0.59_{-0.03}^{+0.03}$ \vspace{2px}\\
$\theta_\mathrm{DM}\ [\mathrm{rad}]$  & Position angle      & $\mathcal{U}$($0,\pi$)    & $2.39_{-0.03}^{+0.02}$  & $2.40_{-0.03}^{+0.02}$ \vspace{2px}\\
$\alpha_\mathrm{DM}$                  & Power-law index     & $\mathcal{U}$($0,2$)      & $1.34_{-0.06}^{+0.04}$  & $1.34_{-0.06}^{+0.04}$ \vspace{2px}\\
$r_\mathrm{e,DM}\ [\arcsec]$          & Einstein radius     & $\mathcal{U}$($0.1,50$)   & $25.2_{-1.4}^{+1.4}$    & $24.8_{-1.5}^{+1.5}$ \vspace{2px}\\
$r_\mathrm{core,DM}\ [\arcsec]$       & Core radius         & $\mathcal{U}$($0.5,20$)   & $4.1_{-0.8}^{+1.1}$     & $3.9_{-0.8}^{+1.1}$ \vspace{2px}\vspace{2px}\\

\midrule

\multicolumn{5}{l}{\textbf{Foreground galaxy} (Spherical dPIE; $z=0.3092$)} \\ 
\midrule
$(x_\mathrm{fg}, y_\mathrm{fg})\ [\arcsec]$  & Centroid            & fixed                    & $(-0.868, -16.564)$  & $(-0.868, -16.564)$ \vspace{2px} \\
$\sigma_\mathrm{fg}\ [\mathrm{km/s}]$        & Velocity dispersion & $\mathcal{U}$($20,300$)  & $79_{-9}^{+24}$      & $80_{-8}^{+17}$ \vspace{2px} \\  
$r_\mathrm{core,fg}\ [\arcsec]$              & Core radius         & fixed                    & $0.001$              & $0.001$ \vspace{2px} \\  
$r_\mathrm{cut,fg}\ [\arcsec]$               & Truncation radius   & $\mathcal{U}$($0.01,30$) & $8.8_{-7.6}^{+14.2}$ & $10.0_{-8.4}^{+13.4}$ \vspace{2px} \\  

\midrule
\multicolumn{5}{l}{\textbf{Background galaxy} (Spherical dPIE; $z=0.3708$)} \\
\midrule
$(x_{\mathrm bg}, y_{\mathrm bg})\ [\arcsec]$ & Centroid            & fixed                       & $(6.871, 3.843)$  & $(6.871, 3.843)$ \vspace{2px} \\
$\sigma_\mathrm{bg}\ [\textrm{km/s}]$         & Velocity dispersion & $\mathcal{N}$($237.7,20.0$) & $231_{-14}^{+14}$ & $232_{-14}^{+14}$ \vspace{2px} \\  
$r_\mathrm{core,bg}\ [\arcsec]$               & Core radius         & fixed                       & $0.001$           & $0.001$ \vspace{2px} \\  
$r_\mathrm{cut,bg}\ [\arcsec]$                & Truncation radius   & $\mathcal{U}$($0.01,50$)    & $30_{-16}^{+14}$  & $30_{-16}^{+14}$ \vspace{2px} \\  

\midrule
\multicolumn{5}{l}{\textbf{Jellyfish galaxies} (Spherical dPIE; $z=z_{\mathrm{l}}$)} \\
\midrule
$(x_\mathrm{JF1}, y_\mathrm{JF1})\ [\arcsec]$ & Centroid            & fixed                    & $(19.094, -13.361)$ & $(19.094, -13.361)$ \vspace{2px} \\
$\sigma_\mathrm{JF1}\ [\textrm{km/s}]$        & Velocity dispersion & $\mathcal{U}$($40,300$)  & $64_{-16}^{+23}$    & $64_{-17}^{+23}$ \vspace{2px} \\
$r_\mathrm{core,JF1}\ [\arcsec]$              & Core radius         & fixed                    & $0.001$             & $0.001$ \vspace{2px} \\  
$r_\mathrm{cut,JF1}\ [\arcsec]$               & Truncation radius   & $\mathcal{U}$($0.01,30$) & $14_{-11}^{+11}$    & $14_{-11}^{+11}$ \vspace{6px} \\

$(x_\mathrm{JF2}, y_\mathrm{JF2})\ [\arcsec]$ & Centroid            & fixed                    & $(-5.114, 6.941)$ & $(-5.114, 6.941)$ \vspace{2px} \\  
$\sigma_\mathrm{JF2}\ [\mathrm{km/s}]$        & Velocity dispersion & $\mathcal{U}$($40,300$)  & $168_{-53}^{+52}$ & $167_{-53}^{+51}$ \vspace{2px} \\
$r_\mathrm{core,JF2}\ [\arcsec]$              & Core radius         & fixed                    & $0.001$           & $0.001$ \vspace{2px} \\  
$r_\mathrm{cut,JF2}\ [\arcsec]$               & Truncation radius   & $\mathcal{U}$($0.01,30$) & $13_{-10}^{+11}$  & $13_{-10}^{+11}$ \vspace{6px} \\

$(x_\mathrm{JF3}, y_\mathrm{JF3})\ [\arcsec]$ & Centroid            & fixed                    & $(-1.649, 2.597)$ & $(-1.649, 2.597)$ \vspace{2px} \\
$\sigma_\mathrm{JF3}\ [\mathrm{km/s}]$        & Velocity dispersion & $\mathcal{U}$($20,300$)  & $56_{-12}^{+35}$  & $56_{-12}^{+36}$ \vspace{2px}\\
$r_\mathrm{core,JF3}\ [\arcsec]$              & Core radius         & fixed                    & $0.001$           & $0.001$ \vspace{2px} \\  
$r_\mathrm{cut,JF3}\ [\arcsec]$               & Truncation radius   & $\mathcal{U}$($0.01,30$) & $7_{-7}^{+15}$    & $7_{-6}^{+15}$ \\  

\midrule
\multicolumn{5}{l}{\textbf{Shear} ($z=z_{\mathrm{l}}$)} \\
\midrule
$\gamma$                        & Shear modulus   & $\mathcal{U}$($0.01,1$) & $0.14_{-0.02}^{+0.02}$ & $0.14_{-0.02}^{+0.02}$ \vspace{2px} \\ 
$\theta_\gamma\ [\mathrm{rad}]$ & Shear direction & $\mathcal{U}$($0, \pi$) & $2.43_{-0.06}^{+0.07}$ & $2.42_{-0.06}^{+0.08}$ \vspace{2px} \\
\bottomrule
\end{tabular}
\label{tab:modelparams}
\caption*{Notes. Columns 1 and 2 list the model parameters and their descriptions. Column 3 reports the adopted priors; the symbol $\mathcal{U}$ denotes a uniform prior, while $\mathcal{N}$ denotes a Gaussian one, with mean and standard deviation given in parentheses. Columns 4 and 5 present the inferred values for the parameters of the corresponding mass models. Parameters reported without uncertainties (``fixed’’) were held fixed in the modelling. The $x$ and $y$ image positions are given in arcseconds relative to the BCG.} 
\end{table}

\newpage

\section{Corner plot}
\label{apdx:C}

\begin{figure*}[ht!]
    \centering
     \includegraphics[width=0.99\textwidth]{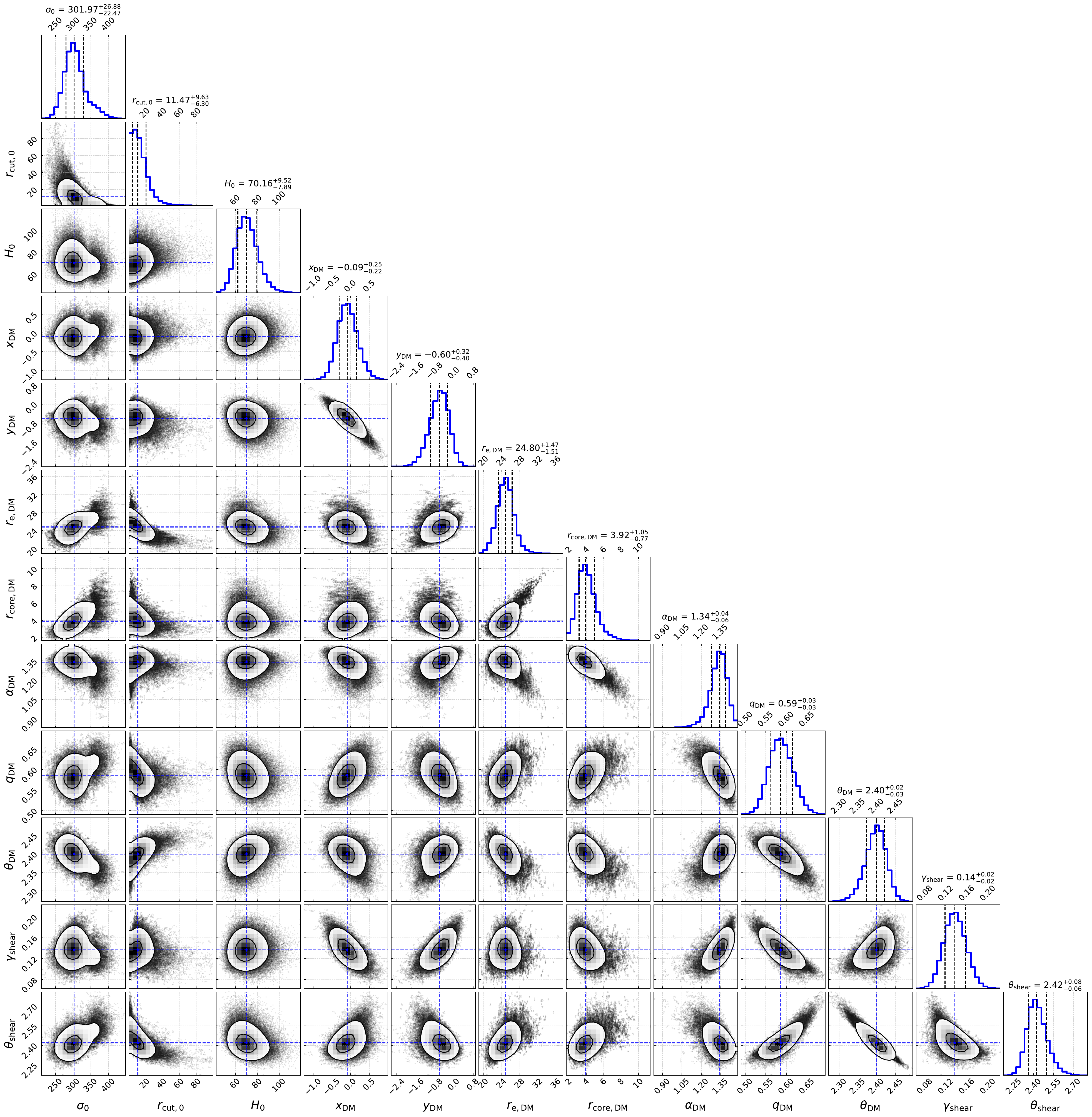}
     \caption{Posterior probability distributions for the Hubble constant and the cluster-scale and sub-halo mass parameters of the \texttt{H0\_free\_SPEMD} lens model. Contours represent $1$ and $2\,\sigma$ confidence levels, while the three vertical dashed lines indicate the $16$th, $50$th, and $84$th percentiles.}
      \label{fig:corner}
\end{figure*}

We present in Fig.~\ref{fig:corner} the posterior probability distributions of the parameters for the cluster-scale and sub-halo mass components and the Hubble constant, for the \texttt{H0\_free\_SPEMD} lens model.

\twocolumn

\end{appendix}

\end{document}